\documentclass[aps,superscriptaddress,twocolumn,nofootinbib,floatfix,prl]{revtex4} 
\pdfoutput=1
\usepackage{graphicx}
\usepackage{bm}

\begin{document}

\title{\boldmath An Improved Standard Model Prediction Of $BR(B \to \tau
  \nu)$\\ And Its Implications For New Physics
   \vspace*{0.2cm}
   \begin{figure}[htb!]
  \begin{center}
  \includegraphics[width=0.13\textwidth]{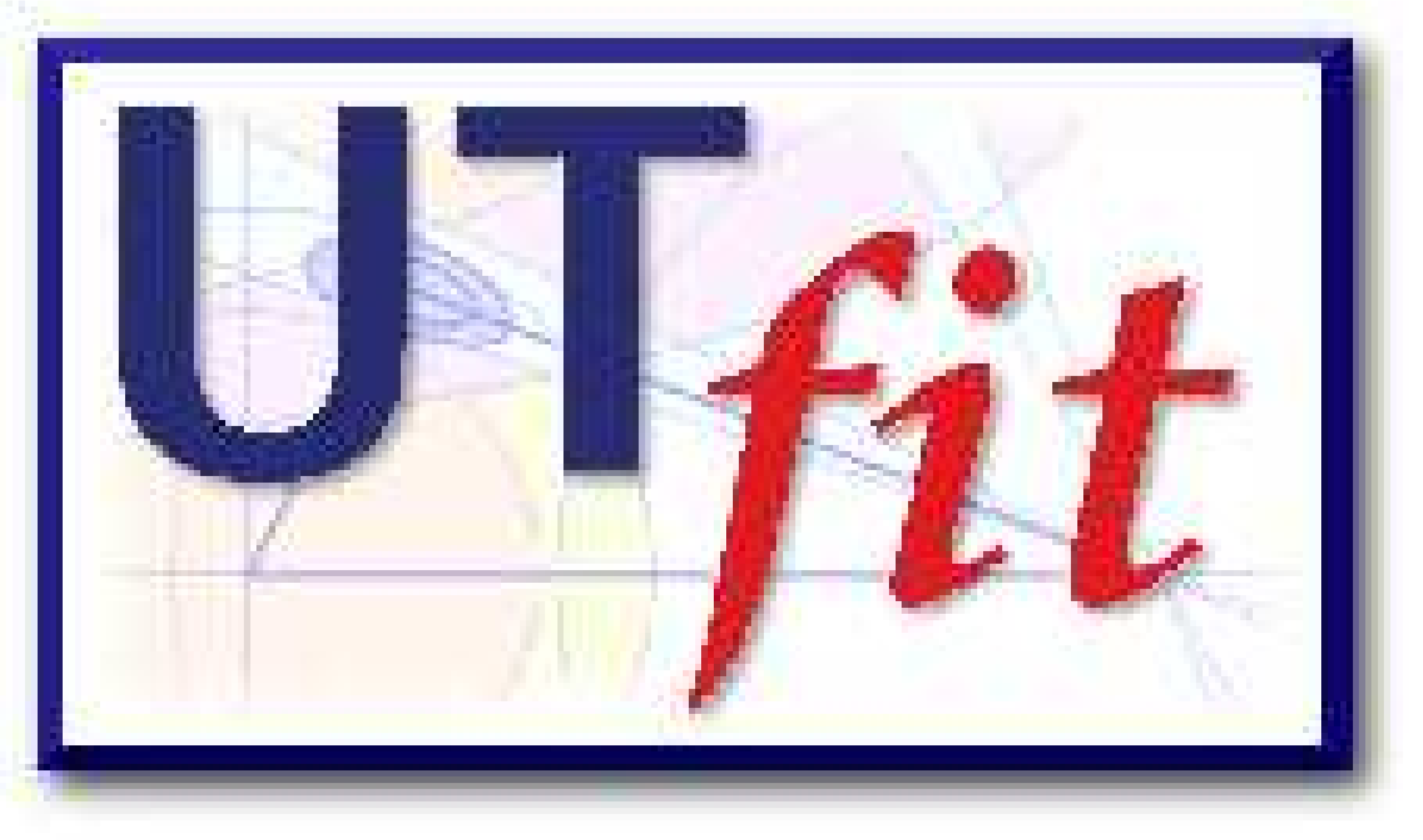}
  \end{center}
 \end{figure}
 \vspace*{-0.8cm}
} 

\collaboration{\textbf{UT}\textit{fit} Collaboration}
\noaffiliation
\author{M.~Bona}
\affiliation{CERN, CH-1211 Geneva 23, Switzerland}
\author{M.~Ciuchini}
\affiliation{INFN,  Sezione di Roma Tre, I-00146 Roma, Italy}
\author{E.~Franco}
\affiliation{INFN, Sezione di Roma, I-00185 Roma, Italy}
\author{V.~Lubicz}
\affiliation{INFN,  Sezione di Roma Tre, I-00146 Roma, Italy}
\affiliation{Dipartimento di Fisica, Universit{\`a} di Roma Tre,
  I-00146 Roma, Italy} 
\author{G.~Martinelli}
\affiliation{INFN, Sezione di Roma, I-00185 Roma, Italy}
\affiliation{Dipartimento di Fisica, Universit\`a di Roma ``La
  Sapienza'', I-00185 Roma, Italy} 
\author{F.~Parodi}
\affiliation{ Dipartimento di Fisica, Universit\`a di Genova and INFN, I-16146
  Genova, Italy} 
\author{M.~Pierini}
\affiliation{CERN, CH-1211 Geneva 23, Switzerland}
\author{C.~Schiavi}
\affiliation{ Dipartimento di Fisica, Universit\`a di Genova and INFN, I-16146
  Genova, Italy} 
\author{L.~Silvestrini}
\affiliation{INFN, Sezione di Roma, I-00185 Roma, Italy}
\author{V.~Sordini}
\affiliation{ETH Zurich, HG Raemistrasse 101, 8092 Zurich, Switzerland}
\author{A.~Stocchi}
\affiliation{Laboratoire de l'Acc\'el\'erateur Lin\'eaire, IN2P3-CNRS et
  Universit\'e de Paris-Sud, BP 34, 
      F-91898 Orsay Cedex, France}
\author{C.~Tarantino}
\affiliation{INFN,  Sezione di Roma Tre, I-00146 Roma, Italy}
\affiliation{Dipartimento di Fisica, Universit{\`a} di Roma Tre,
  I-00146 Roma, Italy} 
\author{V.~Vagnoni}
\affiliation{INFN, Sezione di Bologna,  I-40126 Bologna, Italy}

\begin{abstract}
  The recently measured $B \to \tau \nu$ branching ratio allows to
  test the Standard Model by probing virtual effects of new heavy
  particles, such as a charged Higgs boson. The accuracy of the test
  is currently limited by the experimental error on $BR(B \to \tau
  \nu)$ and by the uncertainty on the parameters $f_B$ and $\vert
  V_{ub}\vert$.  The redundancy of the Unitarity Triangle fit allows
  to reduce the error on these parameters and thus to perform a more
  precise test of the Standard Model.  Using the current experimental
  inputs, we obtain $BR(B \to \tau \nu)_\mathrm{SM} = (0.84 \pm
  0.11)\times 10^{-4}$, to be compared with $BR(B \to \tau
  \nu)_\mathrm{exp} = (1.73 \pm 0.34)\times 10^{-4}$. The Standard
  Model prediction can be modified by New Physics effects in the decay
  amplitude as well as in the Unitarity Triangle fit. We discuss how
  to disentangle the two possible contributions in the case of minimal
  flavour violation at large $\tan\beta$ and generic loop-mediated New
  Physics. We also consider two specific models with minimal flavour
  violation: the Type-II Two Higgs Doublet Model and the Minimal
  Supersymmetric Standard Model.
\end{abstract}

\maketitle

\section{Introduction}
Flavour physics offers the opportunity to probe virtual effects of new
heavy particles using low-energy phenomena, involving Standard Model
(SM) particles as external states. New Physics
(NP) can generate large effects in Flavour Changing Neutral Currents
(FCNC) and CP violating phenomena even for NP particle masses much
above the TeV scale, if new sources of flavour and CP violation
besides the Yukawa couplings are present. The strong NP sensitivity is
mainly due to the Glashow-Iliopoulos-Maiani (GIM) suppression of FCNC
processes in the SM~\cite{GIM}. However, other suppression mechanisms
can be at work in the SM, making a few non-FCNC decays interesting for
NP searches. In particular, the helicity suppression of the charged
current decay $B \to \tau \nu$ makes it potentially sensitive to the
tree-level effects of new scalar particles~\cite{btaunujurassic}. A
typical example is given by the exchange of charged Higgs bosons in
multi-Higgs extensions of the SM, such as the type-II Two Higgs Doublet
Model (2HDM-II) or the Minimal Supersymmetric Standard Model (MSSM),
in the large $\tan\beta$ regime.

In the SM, the branching ratio of $B \to \tau \nu$ can be written
as:
\begin{equation}
BR(B \to \tau \nu) = 
\frac{G_{F}^{2}m_{B}m_{\tau}^{2}}{8\pi}\left(1-\frac{m_{\tau}^{2}}
{m_{B}^{2}}\right)^{2}f_{B}^{2}|V_{ub}|^{2}\tau_{B}\,.
\label{eq:btaunu}
\end{equation}
The Fermi constant $G_F$, the $B$ ($\tau$) mass $m_B$ ($m_\tau$) and
the $B$ lifetime $\tau_B$ are precisely measured~\cite{PDG}.  The
decay constant of the $B$ meson $f_B$ is known with ${\cal O}(10\%)$
uncertainty. We use the lattice QCD (LQCD) average $f_B = 200 \pm 20$
MeV~\cite{latticefB}. Concerning the error attached to lattice
averages, we combine in quadrature the statistical and systematic
errors, assuming Gaussian distributions.  This is justified since
present lattice systematic errors arise from the combination of
several independent sources of uncertainty. Therefore they are well
described by a Gaussian distribution, no matter what the distributions
of the individual sources are.~\footnote{Notice that in the past we
  used to assign a flat distribution to the lattice systematic errors,
  since they were dominated by the uncertainty associated to the
  quenched approximation.}

The absolute value of the Cabibbo-Kobayashi-Maskawa (CKM)~\cite{ckm}
matrix element $V_{ub}$ is determined from the measurements of the
branching ratios of {\it exclusive} and {\it inclusive} semileptonic
$b \to u$ decays. Its precision is limited by the uncertainty of the
theoretical calculations. Although inclusive determinations are
systematically higher than exclusive ones, the two values are
compatible, once the spread of inclusive determinations using
different theoretical models is considered. For the exclusive decays,
we use the HFAG averages~\cite{bpilnu,HFAG}
\begin{eqnarray}
&&\hspace{-0.4cm}BR(B\to\pi\ell\nu)_{q^2<16\,\mathrm{GeV}^2}=(0.94\pm
0.05\pm 0.04)\times 
10^{-4}\,,\nonumber\\
&&\hspace{-0.4cm}BR(B\to\pi\ell\nu)_{q^2>16\,\mathrm{GeV}^2}=(0.37\pm
0.03\pm 0.02)\times 10^{-4}\,,\nonumber
\end{eqnarray}
together with the theoretical estimates of the relevant normalized
form factors
\begin{center}
  \begin{tabular}{c}
    $FF(q^2<16\,\mathrm{GeV}^2)=5.44\pm 1.43$~\cite{ballzwicky}\,,\\
    $FF(q^2>16\,\mathrm{GeV}^2)=2.04\pm 0.40$~\cite{latticefB}\,,
  \end{tabular}\\
\end{center}
to obtain $\vert V_{ub}\vert^\mathrm{excl}=(33.3\pm 2.7)\times 10^{-4}$.
For inclusive decays, we quote $\vert V_{ub}\vert^\mathrm{incl}=(40.0\pm 1.5\pm
4.0)\times 10^{-4}$, where we define the second error as a flat range
accounting for the spread of the different models~\cite{vubinclth}.

\begin{figure}[tb]
    \includegraphics[width=0.38\textwidth]{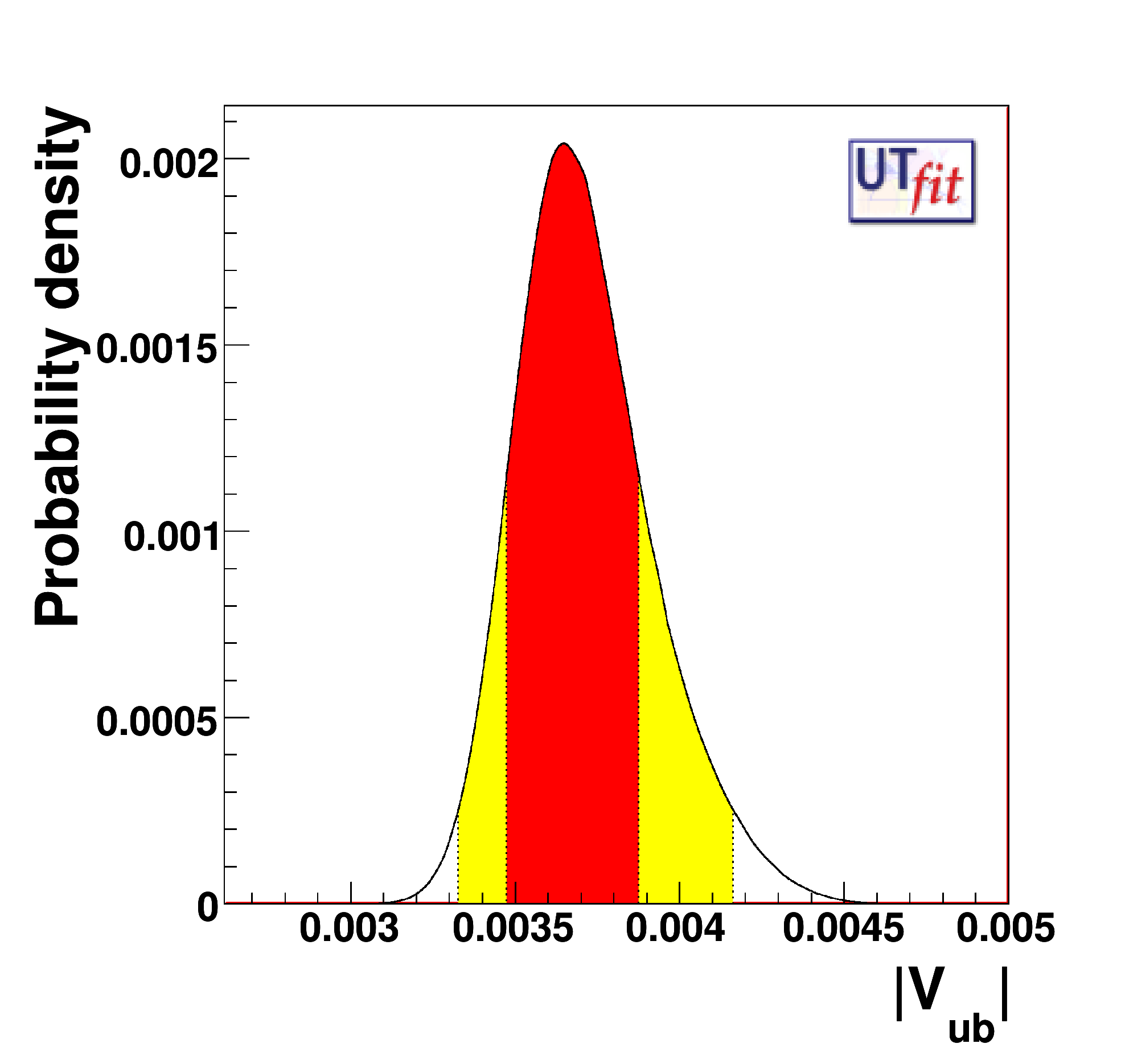} 
    \caption{P.d.f.\ of $\vert V_{ub}\vert$ obtained combining
    inclusive and exclusive measurements of the 
    $b\to u$ semileptonic decays. The dark
      (light) region corresponds to the $68\%$ ($95\%$) probability
      interval.\label{fig:vubinp}}
\end{figure}

Our grand average of inclusive and exclusive determinations is $\vert
V_{ub}\vert=(36.7\pm 2.1) \times 10^{-4}$, obtained from the
probability density function (p.d.f.) in Fig.~\ref{fig:vubinp}.  From
this p.d.f.\ we get
\begin{equation}
BR(B \to \tau \nu) = (0.98 \pm 0.24)\times 10^{-4}\,, \label{eq:nofit} 
\end{equation}
compatible with $BR_\mathrm{exp} = (1.73 \pm
0.34)\times 10^{-4}$ \cite{btaunuexp} at $\sim 1.8\sigma$.

A few percent precision is expected to be reached by LQCD using
Petaflop CPUs for $f_B$ and the form factors entering the exclusive
determination of $\vert V_{ub}\vert$~\cite{superBcdr}.  Considering
how challenging the measurement of $BR(B \to \tau \nu)$ in a hadronic
environment is, it is difficult to imagine a similar improvement in
precision of the experimental measurement, unless a SuperB factory
will be built, leading also to a better direct determination of $\vert
V_{ub}\vert$~\cite{superBcdr}.
On the other hand, it has been pointed out in Ref.~\cite{utfitBtaunu}
that the indirect
determination of $\vert V_{ub}\vert$ from the Unitarity Triangle (UT)
fit in the SM is more accurate than the measurements, yielding a
central value close to the {\it exclusive} determination.  Therefore a
more precise prediction of $BR(B \to \tau \nu)$ in the SM can be
obtained combining the {\it direct} knowledge of $|V_{ub}|$ and $f_B$
with the {\it indirect} determination from the rest of the UT fit.

\section{UTfit-improved Standard Model prediction}

In the UT fit~\cite{utfit,ckmfitter}, CP-conserving and CP-violating
measurements are combined to constrain $\bar\rho$ and $\bar\eta$. The
fit also provides an {\it a-posteriori} determination of $\vert
V_{ub}\vert$ which includes the {\it direct} measurement as well as
the {\it indirect} determination from the other constraints.
Similarly, an improved determination of $f_B$ from both LQCD and
experimental constraints is obtained~\cite{utfitBtaunu}.

\begin{figure}[tb]
    \includegraphics[width=0.38\textwidth]{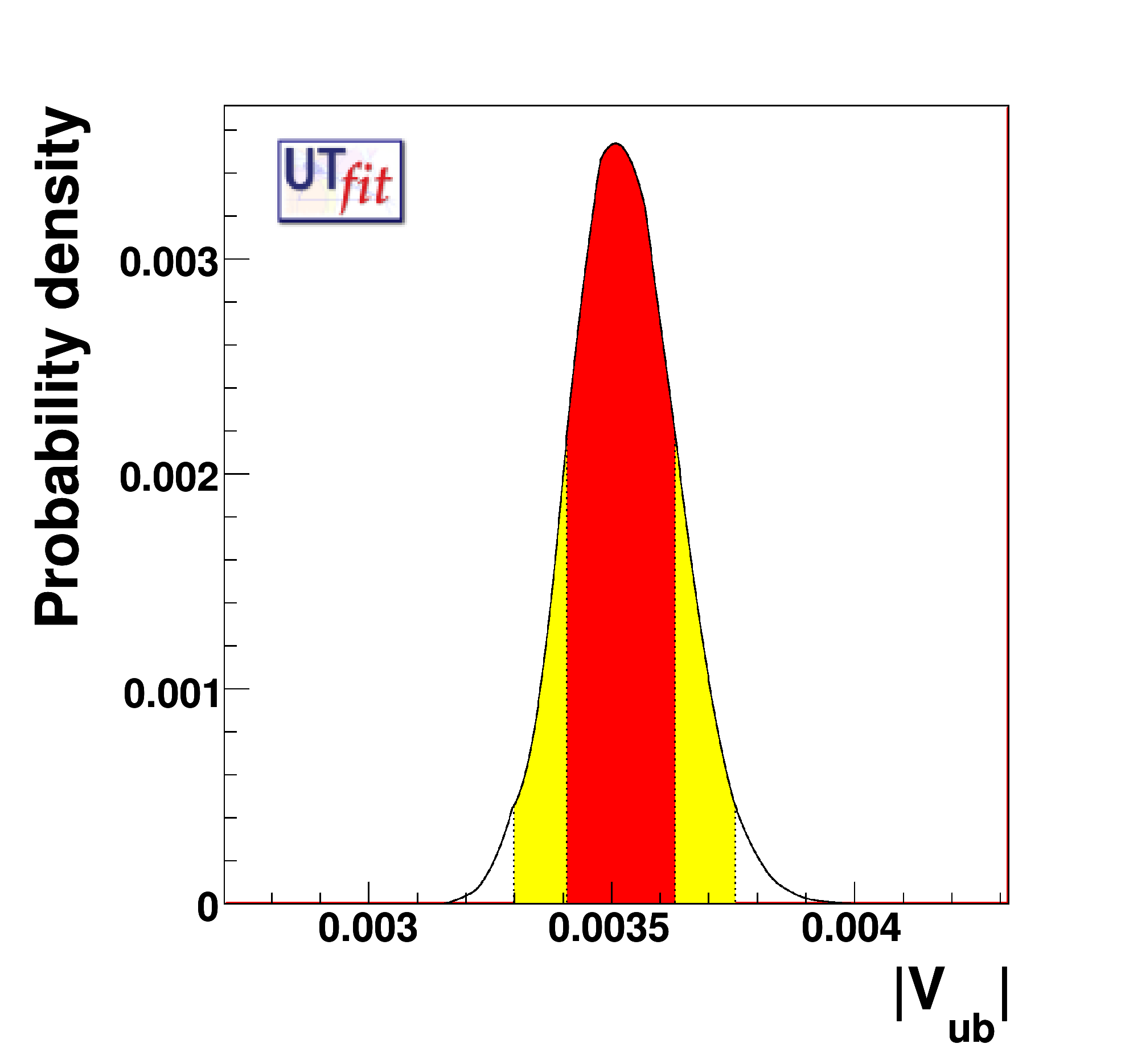} \\ 
    \includegraphics[width=0.38\textwidth]{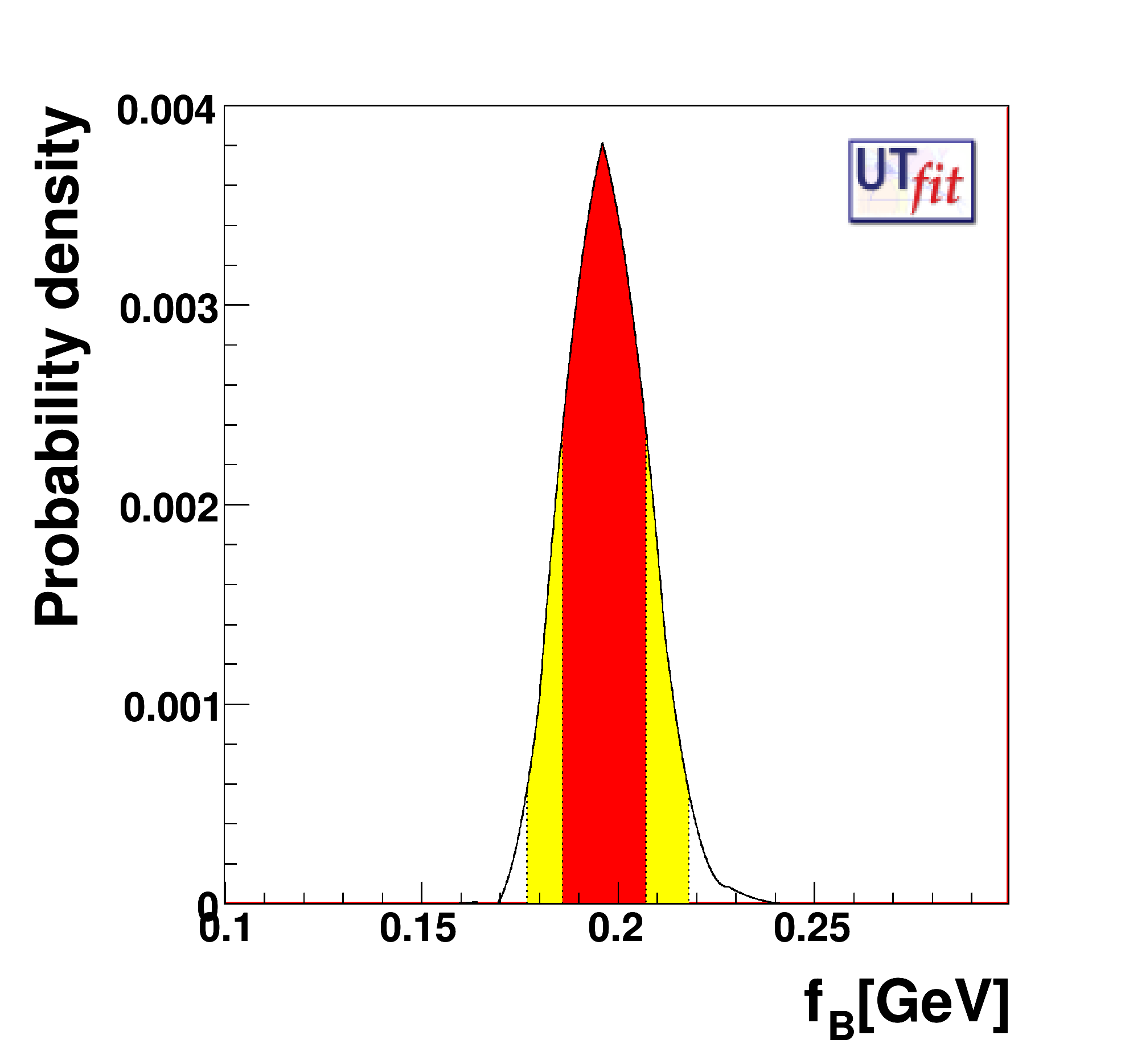} 
    \caption{Posterior p.d.f.\ for $\vert V_{ub}\vert$ (top) and $f_B$ (bottom),
      obtained from the UT fit, without taking $BR(B \to \tau \nu)$ as
      input.  The dark
      (light) region corresponds to the $68\%$ ($95\%$) probability
      interval.\label{fig:vubfB}}
\end{figure}

The most accurate prediction of $BR(B \to \tau \nu)$ in the SM can
then be obtained performing the SM fit without including the
measurement of $BR(B \to \tau \nu)$ as a constraint.  The fit gives
$\bar \rho = 0.149 \pm 0.021$ and $\bar \eta = 0.334 \pm 0.013$
together with $f_B = (196 \pm 11)$ MeV and $\vert V_{ub}\vert = (35.2
\pm 1.1)\times 10^{-4}$. The posterior p.d.f.'s are shown in
Fig.~\ref{fig:vubfB}.

\begin{figure}[tb]	
    \includegraphics[width=0.38\textwidth]{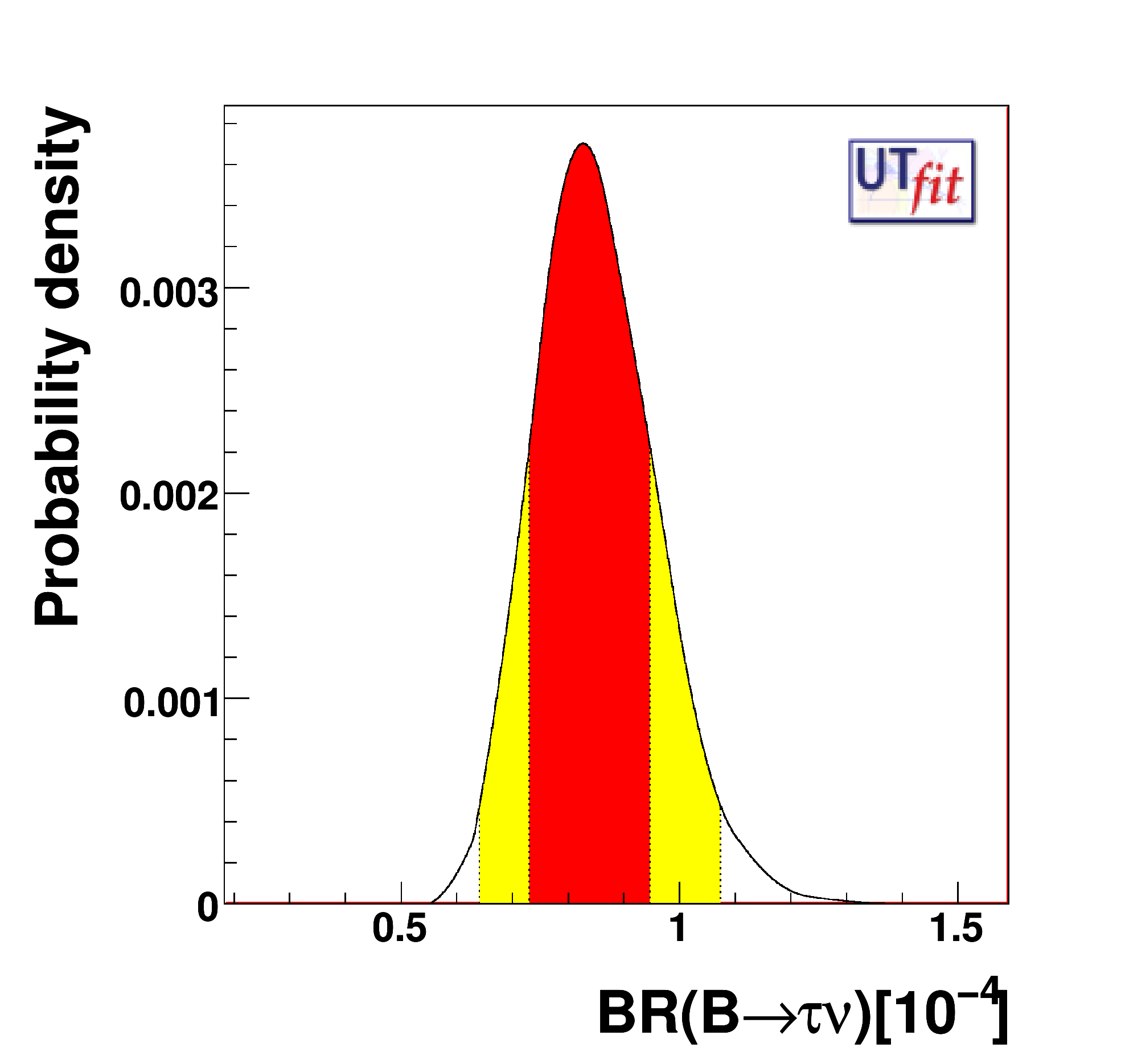} \\
    \caption{P.d.f.\ for $BR(B\to\tau\nu)$ predicted using the UT fit.
      The dark (light) region corresponds to the $68\%$ ($95\%$)
      probability interval.}
\label{fig:btaunuratio}
\end{figure}

\begin{figure}[tb]	
    \includegraphics[width=0.38\textwidth]{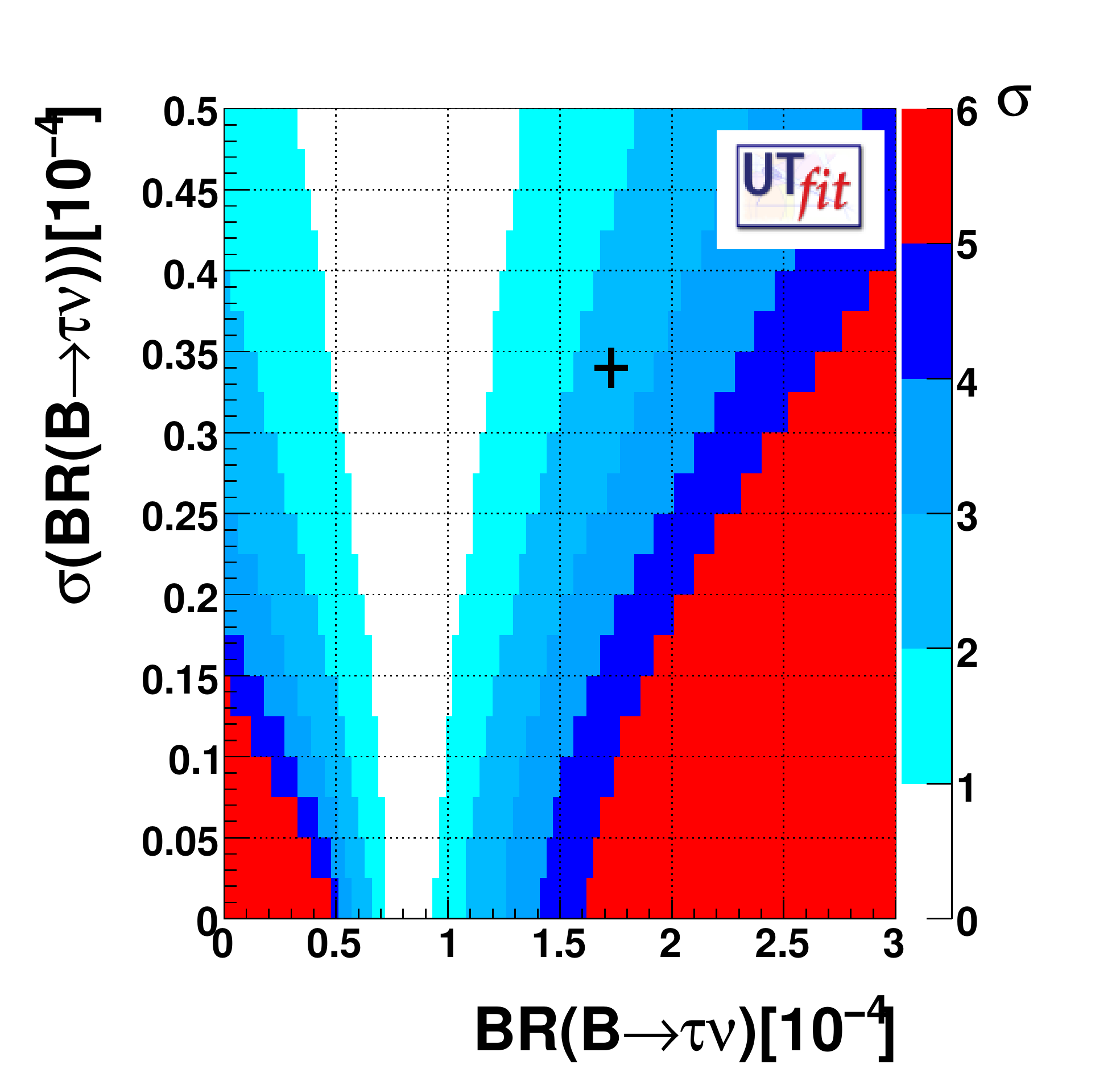} 
    \caption{Compatibility plot for $BR(B \to \tau
      \nu)$. The cross marks the current world average; colours give
      the agreement (in number of $\sigma$) with the data-driven SM
      prediction.}
\label{fig:comp}
\end{figure}

The same SM fit gives the p.d.f.\ in Fig.~\ref{fig:btaunuratio}, from
which we obtain
\begin{equation}
BR(B \to \tau \nu)_\mathrm{SM} = (0.84 \pm 0.11) \times 10^{-4}\,.
\label{btaunuUTprediction}
\end{equation}
In Fig.~\ref{fig:comp} we present the compatibility plot for $BR(B \to
\tau \nu)_\mathrm{SM}$. The colored regions represent the pull from
the UT fit result. The present experimental value, represented by a
cross in the plot, displays a deviation of $\sim 2.5\sigma$. This
deviation can be interpreted as a similar same-sign statistical
fluctuation (or a correlated systematic error) in BaBar and Belle
results or as a hint of NP effects. A more definite answer needs new
data to be collected.

From Eq.~(\ref{btaunuUTprediction}), one can easily predict
the SM value of $BR(B \to \mu \nu)$ and $BR(B \to e \nu)$. We obtain
\begin{eqnarray}
&&BR(B\to \mu\nu)_\mathrm{SM} = (3.8 \pm 0.5) \times 10^{-7}\,, \\
&&BR(B\to e\nu)_\mathrm{SM} = (8.8 \pm 1.2)\times  10^{-12}\,. \nonumber
\end{eqnarray}
The precision on the experimental measurements~\cite{expOther} is
still far from probing such small values. The current best limits are
$BR(B\to \mu \nu)<1.0 \times 10^{-6}$~\cite{bmunubabarnew} and $BR(B
\to e \nu)<1.0 \times 10^{-6}$~\cite{HFAG} at $90\%$ C.L.

\section{Model-independent predictions}
Let us assume in the following that NP is at work. In this case, the
prediction in Eq.~(\ref{btaunuUTprediction}) could be modified by
\textit{i)} NP effects in the decay amplitude and/or \textit{ii)} NP
effects in the UT fit.  If more precise measurements will provide
evidence of a discrepancy, one should be careful in interpreting it as
evidence of NP in the $B \to \tau \nu$ decay amplitude. In fact, other
inputs of the UT analysis (for example $\Delta m_q$ ($q=d,s$)) might
be affected by the presence of contributions beyond the SM.
We would like to disentangle the two possible NP effects.
To this aim, we compute the
prediction of $BR(B \to \tau \nu)$ in several NP scenarios
\textit{assuming} that NP contributions to the $B \to \tau \nu$ decay
amplitude are negligible. This prediction will be denoted as
$\overline{BR}_\mathrm{model}$. A discrepancy between
$\overline{BR}_\mathrm{model}$ and $BR_\mathrm{exp}$ would
unambiguously reveal NP contributions to the $B \to \tau \nu$ decay
amplitude in the considered scenario.

As is common practice in the literature, we also provide results in
terms of the ratio
\begin{equation}
R_\mathrm{model}^\mathrm{exp} =
\frac{BR_\mathrm{exp}}{\overline{BR}_\mathrm{model}}\,. 
\label{eq:rdef}
\end{equation}
The use of $R^\mathrm{exp}_\mathrm{model}$ is particularly convenient
for NP models with Minimal Flavour Violation (MFV)~\cite{MFV,amb},
defined as models where the only source of flavour violation are the
quark masses and the CKM matrix~\cite{amb}.
Indeed, $BR_\mathrm{MFV}$, the full prediction of the branching ratio
including NP in the decay amplitude, and $\overline{BR}_\mathrm{MFV}$
have the same dependence on $\vert V_{ub}\vert$ and $f_B$, so that they
cancel in the ratio $R_\mathrm{MFV}=BR_\mathrm{MFV}/\overline{BR}_\mathrm{MFV}$.
Therefore, $R_\mathrm{MFV}$ can be computed theoretically without specifying
the value of $\vert V_{ub}\vert$ and $f_B$. $R_\mathrm{MFV}$ is constrained by
$R_\mathrm{MFV}^\mathrm{exp}$, which contains the experimental error
as well as the uncertainty on $\vert V_{ub}\vert$ and $f_B$.

Following Ref.~\cite{UTDF2np}, we distinguish several scenarios
according to the NP flavor structure. In each scenario, we remove all
the inputs that might be affected by NP from the
UTfit-based determination of $BR(B \to \tau\nu)$. This
gives a NP-independent prediction of $\overline{BR}$.

\begin{figure}[tb]	
    \includegraphics[width=0.38\textwidth]{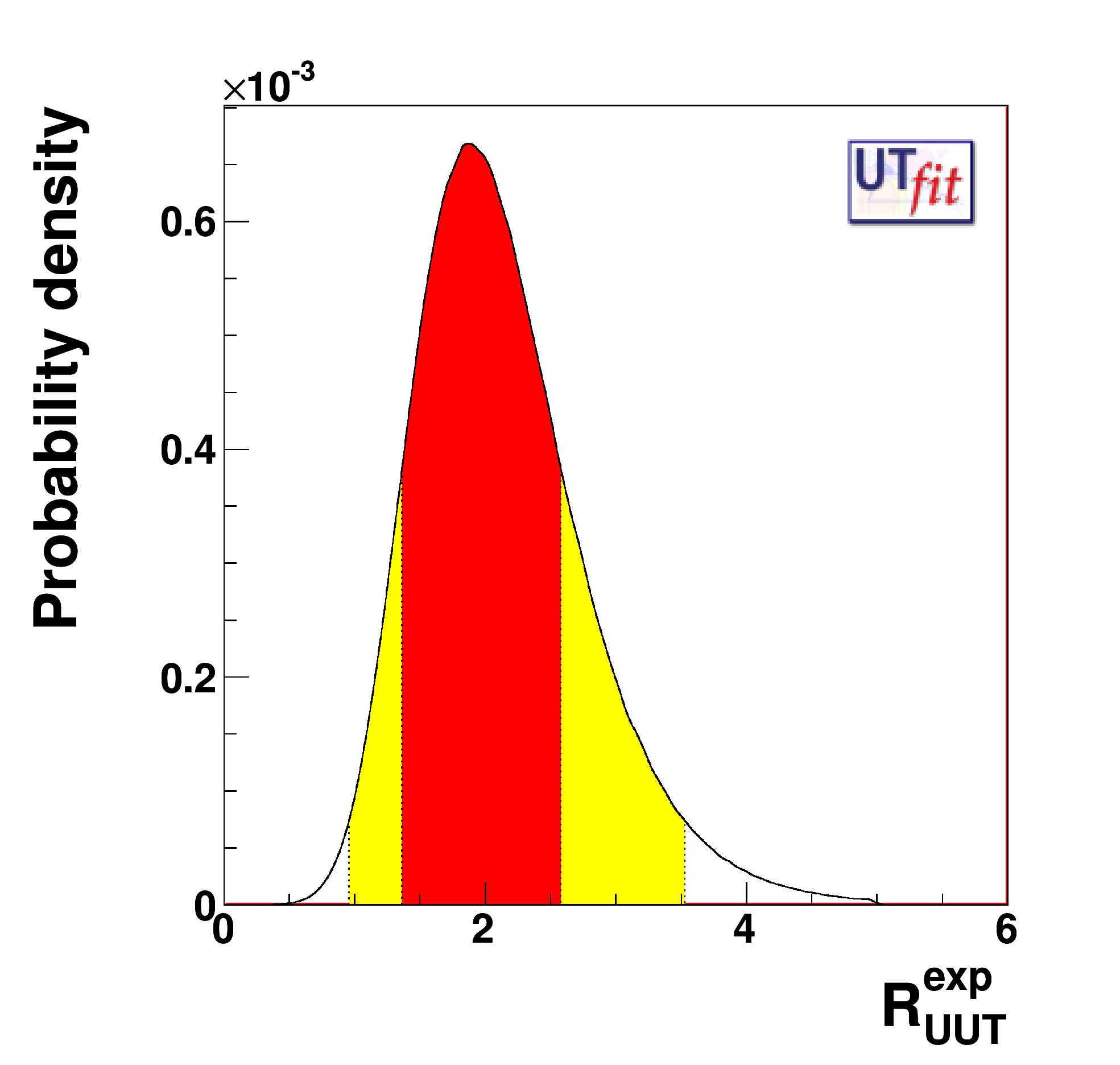}
    \caption{P.d.f.\ of $R_\mathrm{UUT}^\mathrm{exp}$ obtained using the UUT
      construction.\label{fig:RUUT}}
\end{figure}

In MFV models one expects the tree-level processes and the angles of
the UT not to deviate from the SM prediction, while the values of
$\Delta m_q$ and $\epsilon_K$ are expected to change.\footnote{In MFV models
  one has to assume that the large measured value of the $B_s$
  mixing phase is a statistical fluctuation. Otherwise, MFV would be
  excluded~\cite{phibs}.} We can then replace the full SM UT fit with
the Universal UT (UUT) construction~\cite{UUT}. In the case of the
UUT, the knowledge of $f_B$ is given by LQCD only, resulting in a
larger error on $\overline{BR}_\mathrm{UUT}$. Using the currently
available experimental inputs, we obtain
$\overline{BR}_\mathrm{UUT}=(0.87 \pm 0.20)\times 10^{-4}$
corresponding to $R_\mathrm{UUT}^\mathrm{exp} = 2.0 \pm 0.6$, as shown
in Fig.~\ref{fig:RUUT} (for comparison, see the SM result in
Eq.~(\ref{btaunuUTprediction})). Clearly, the determination of
$\overline{BR}_\mathrm{UUT}$ will benefit considerably from the
expected improvements in future LQCD calculations.

In MFV models with one Higgs doublet (or two Higgs doublets
at small $\tan\beta$), one expects negligible NP effects in the
$B \to \tau \nu$ decay amplitude, while a deviation could
be induced on $\Delta m_d$, $\Delta m_s$, and $\epsilon_K$.
Should $R_\mathrm{UUT}^\mathrm{exp}$ deviate from one significantly,
these models would then be excluded.

\begin{figure*}[tb]	
    \includegraphics[width=0.45\textwidth]{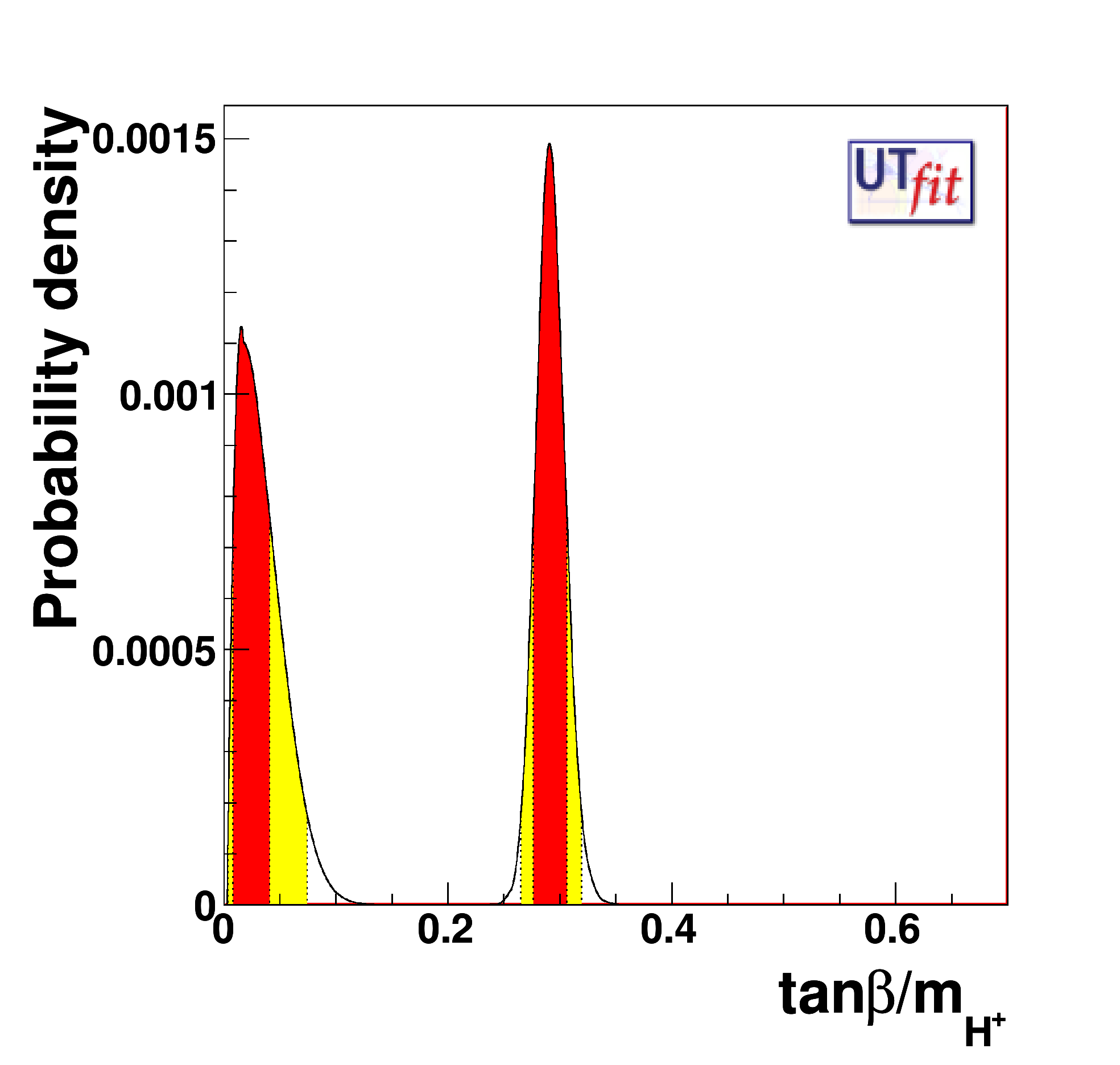} 
    \includegraphics[width=0.45\textwidth]{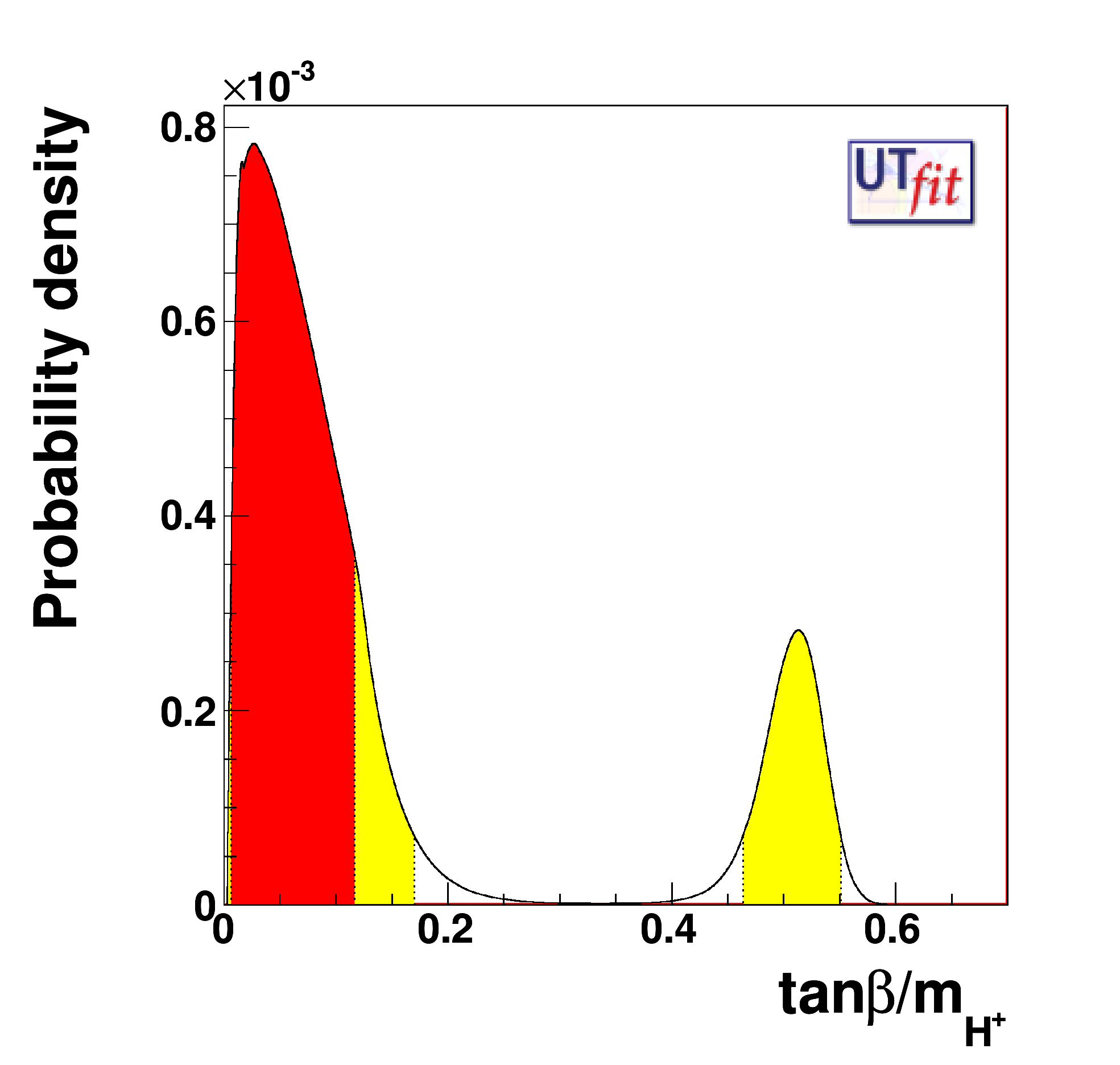} 
    \caption{P.d.f.\ of $\tan\beta/m_{H^+}$ computed from
      Eq.~(\ref{eq:R2hdm}) and the fit result for
      $R_\mathrm{UUT}^\mathrm{exp}$ (left), or using $BR(B
      \to D \tau \nu)/BR(B \to D \ell \nu)$ (right).\label{fig:2HDM}}
\end{figure*}

In the case of MFV models with two Higgs doublets at large $\tan\beta$,
the value of $R_\mathrm{UUT}^\mathrm{exp}$ could be shifted from one by the
contribution of the charged Higgs boson to the decay amplitude.

\section{Constraints on 2HDM-II}
\label{sec:2HDMII}

As an explicit example of the discussion above, we consider the
2HDM-II. In this model, the interaction between quarks and the
charged Higgs $H^\pm$ is defined by the Lagrangian
\begin{eqnarray}
  {\cal L}&=& (2\sqrt{2}G_F)^{1/2}\sum_{i,j=1}^3\bar u_i
  \Bigl( \frac{1}{\tan\beta} m_{u_i}V_{ij}\frac{1-\gamma_5}{2}\nonumber\\
  &&+\tan\beta\, V_{ij}\, m_{d_i}\frac{1+\gamma_5}{2}\Bigr) d_j H^{+} + \mathrm{H.c.}\,,
\end{eqnarray}
and FCNC are absent at the tree level.

\begin{figure*}[tb]
     \includegraphics[width=0.8\textwidth]{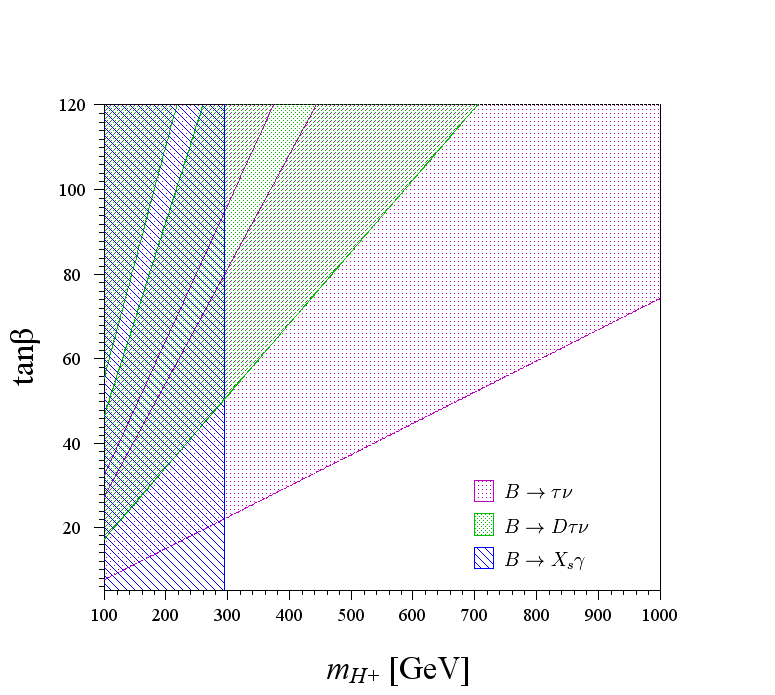} 
     \caption{Regions in the $(m_{H^+},\tan\beta)$ parameter space of
       the 2HDM-II excluded at $95\%$ probability by
       $BR(B\to\tau\nu)$, $BR(B\to D \tau\nu)/BR(B\to D \ell\nu)$ and
       $BR(B\to X_s\gamma)$.\label{fig:tanbmh2d}}
\end{figure*}

We can write~\cite{hou}:
\begin{equation}
R_\mathrm{2HDM} = \left(1-\tan^2\beta \frac{m_B^2}{m_{H^+}^2}\right)^2\,,
\label{eq:R2hdm}
\end{equation}
where $m_{H^+}$ is the mass of the charged Higgs
boson. Eq.~(\ref{eq:R2hdm}), together with the p.d.f.\ of
$R_\mathrm{UUT}^\mathrm{exp}$ provided by the UUT fit, gives a
constraint on $\tan\beta/m_{H^+}$ as shown in Fig.~\ref{fig:2HDM}.
The charged Higgs contribution typically suppresses $BR(B \to \tau
\nu)$ with respect to the SM, contrary to current experimental
results. An excess can be obtained if $\tan\beta > \sqrt{2}\,
m_{H^+}/m_B$ (corresponding to the rightmost peak in the left plot of
Fig.~\ref{fig:2HDM}, $\tan\beta = (29 \pm 2)\, m_{H^+}/(100$ GeV$)$),
yielding an upper limit on $m_{H^+}$ for a given value of $\tan\beta$.
The current direct searches~\cite{directHplus} give a lower limit of
$m_{H^+}>79$ GeV at $95\%$ C.L.~\cite{PDG}, while the measurement of
$BR(B \to X_s \gamma)$ implies $m_{H^+}>295$ GeV at $95\%$ C.L. for
the 2HDM-II charged Higgs boson~\cite{gambino}. This bound excludes
the rightmost peak in Fig.~\ref{fig:2HDM} for $\tan \beta<80$.  In
addition, one can consider the bound on $\tan\beta/m_{H^+}$ from $BR(B
\to D \tau \nu)/BR(B \to D \ell \nu)$ where $\ell$ denotes light
leptons~\cite{btaunujurassic,itoh}. Using the world average $(49\pm
10)\%$~\cite{bdtaunuexp} and formula (9) of Ref.~\cite{mescia} we
obtain the following $95\%$ probability regions for
$\tan\beta/m_{H^+}$: $\tan\beta/m_{H^+}<0.17\, \mathrm{GeV}^{-1}$ and
$0.46\, \mathrm{GeV}^-1 <\tan\beta/m_{H^+}<0.55 \,\mathrm{GeV}^{-1}$
(see the right plot in Fig.~\ref{fig:2HDM}. In this case, as for the
$B \to \tau \nu$ bound, there is an allowed region at large
$\tan\beta/m_{H^+}$.  Assuming flat priors in $[5,120]$ for $\tan
\beta$~\cite{tanbmax} and $[100,1000]$ GeV for $m_{H^+}$, we obtain
the plot in Fig.~\ref{fig:tanbmh2d}. For $\tan\beta\gtrsim 22$
$B\to\tau\nu$ gives a lower bound on $m_{H^+}$ stronger than the one
from $B \to X_s \gamma$. The fine-tuned regions for large
$\tan\beta/m_{H^+}$ allowed individually by the $B \to \tau \nu$ and
the $B \to D \tau \nu$ constraints do not overlap and are therefore
excluded.  We thus obtain an absolute bound
\begin{equation}
  \label{eq:2HDMbound}
  \tan\beta<7.4 \frac{m_{H^+}}{100\, \mathrm{GeV}}\,. 
\end{equation}
In addition, we compute the prediction for $BR(B_s \to \mu^+ \mu^-)$
and obtain
\begin{eqnarray}
\label{eq:bsmm2hdm}
BR(B_s \to \mu^+ \mu^-)&=&(4.3 \pm 0.9)\times 10^{-9} \\
&&([2.5,6.2]\times 10^{-9}~@95\%~\mathrm{prob.}). \nonumber
\end{eqnarray}

Our results are in agreement with Ref.~\cite{0907.5135}, where the
effect of $BR(B \to \tau \nu)$ and other constraints on the 2HDM-II
has been recently analysed. However, our analysis differs in several
aspects. First, in the UUT analysis we keep all the angles, which are
unaffected by MFV NP effects.\footnote{In the extraction of $\sin 2
  \beta$ from $B \to J/\Psi K$ decays, possible NP enhancements of the
  penguin amplitude are bound using additional data \cite{CPS}.}
Second, we neglect sub-percent contributions to tree-level decays,
allowing us to use all determinations of $\vert V_{ub}\vert$. Third,
we only consider the dominant constraints from $B \to X_s \gamma$ and
$B \to \tau \nu$. Finally, we use the Bayesian approach detailed in
Ref.~\cite{dagos}.

One of the most interesting features of the relation in
Eq.~(\ref{eq:R2hdm}) is that it does not depend on the flavour of the
final lepton~\cite{hou}, since the helicity suppression in the SM
compensates the scaling of the Higgs couplings with the mass. This
means that, provided the evidence of a discrepancy in $B\to\tau\nu$,
the same effect should be observed in $B\to\ell\nu$ ($\ell = e,
\mu$). For these decays, we get
\begin{eqnarray}
&& \overline{BR}(B\to \mu\nu)_\mathrm{UUT} = (3.9 \pm 0.9) \times 10^{-7}\,, \\
&& \overline{BR}(B\to e\nu)_\mathrm{UUT} = (9.2 \pm 2.1)\times
10^{-12}\,, \nonumber 
\end{eqnarray}
where $\overline{BR}$ for these decays is defined in analogy with the
$B\to\tau\nu$ case.

Beyond MFV, the UUT construction is no longer adequate. Indeed, in the
most general case, assuming only that NP contributions to semileptonic
decays are negligible, the prediction of $BR(B\to \tau\nu)$ cannot be
improved using the UT fit and the result can be read from
Eq.~(\ref{eq:nofit}), $\overline{BR}_\mathrm{no-fit}=(0.98\pm
0.24)\times 10^{-4}$.

To summarize our results, we collect in Table \ref{tab:results} our
predictions for $\overline{BR}$ in the considered scenarios.

\begin{table}
 \begin{center}
\begin{tabular}{l|r|r|r|r}
 scenario & $\vert V_{ub}\vert\times 10^4$ & $f_B$ (MeV) &
 $\overline{BR}\times 10^4$ & pull\\\hline 
 UT     & $35.2\pm 1.1$ & $196\pm 11$ & $0.84\pm 0.11$ & $2.5\sigma$\\
 UUT    & $35.0\pm 1.2$ & $200\pm 20$ & $0.87\pm 0.20$ & $2.2\sigma$\\
 no-fit & $36.7\pm 2.1$ & $200\pm 20$ & $0.98\pm 0.24$ & $1.8\sigma$
 \end{tabular}
 \end{center}
\caption{Results for $\vert V_{ub}\vert$, $f_B$, $\overline{BR}$ and
  the pull between 
$\overline{BR}$ and $BR(B\to\tau\nu)_\mathrm{exp}$ in different
scenarios (see text).}\label{tab:results}
\end{table}

\section{Constraints on the MSSM parameters}

\begin{figure}[htb]
 \centering
 \includegraphics[width=0.22\textwidth]{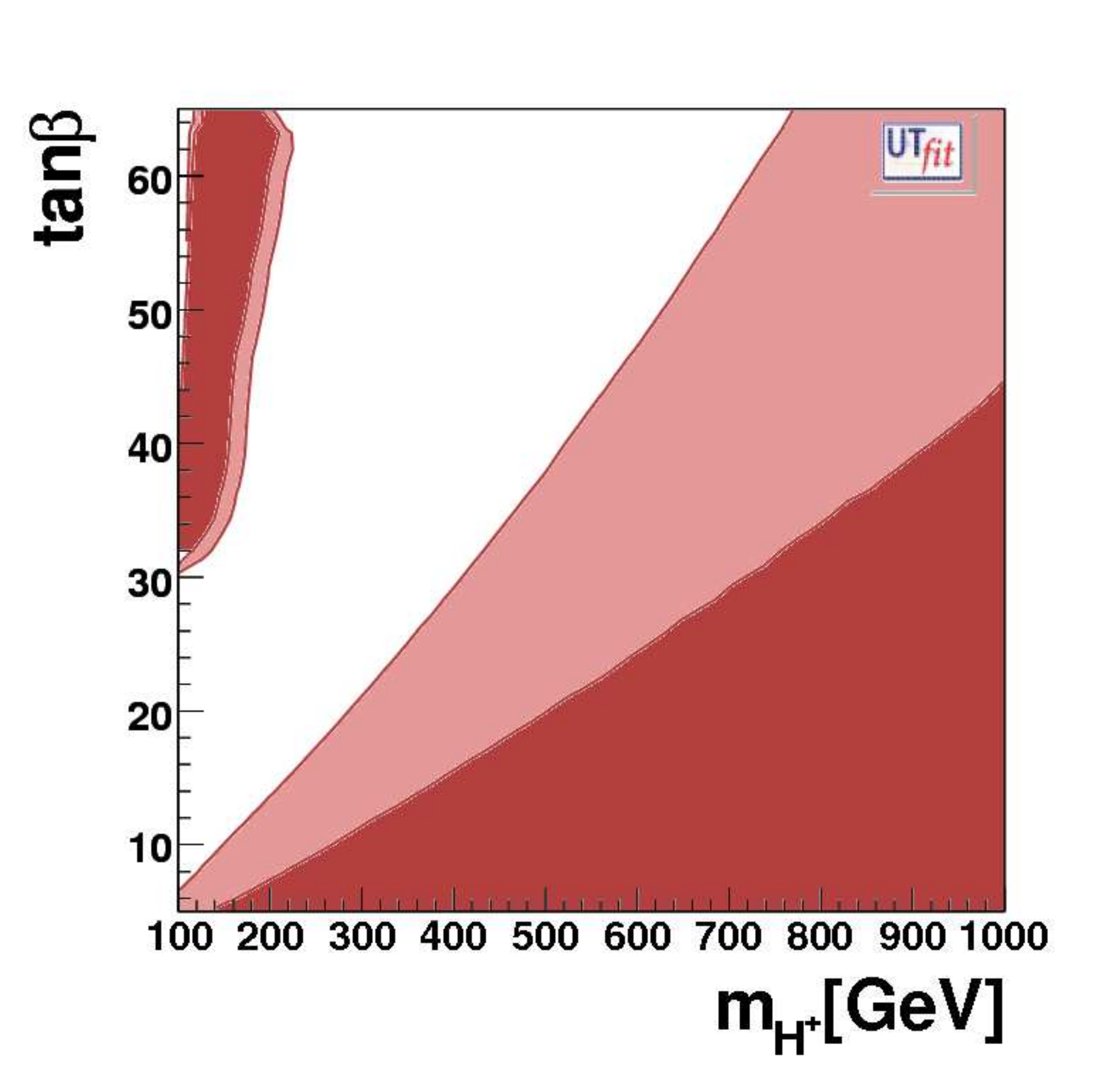}
 \includegraphics[width=0.22\textwidth]{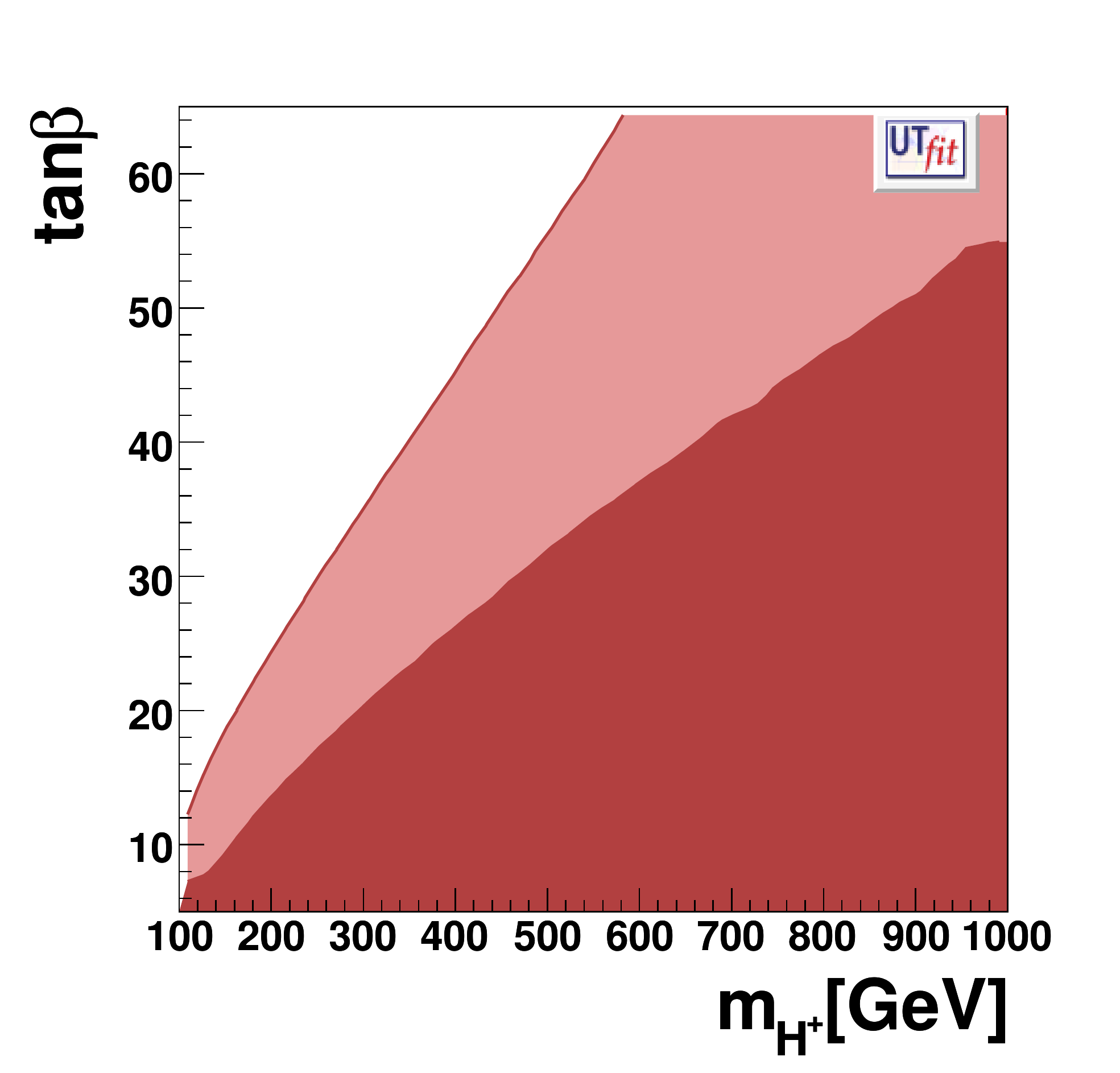}
 \includegraphics[width=0.22\textwidth]{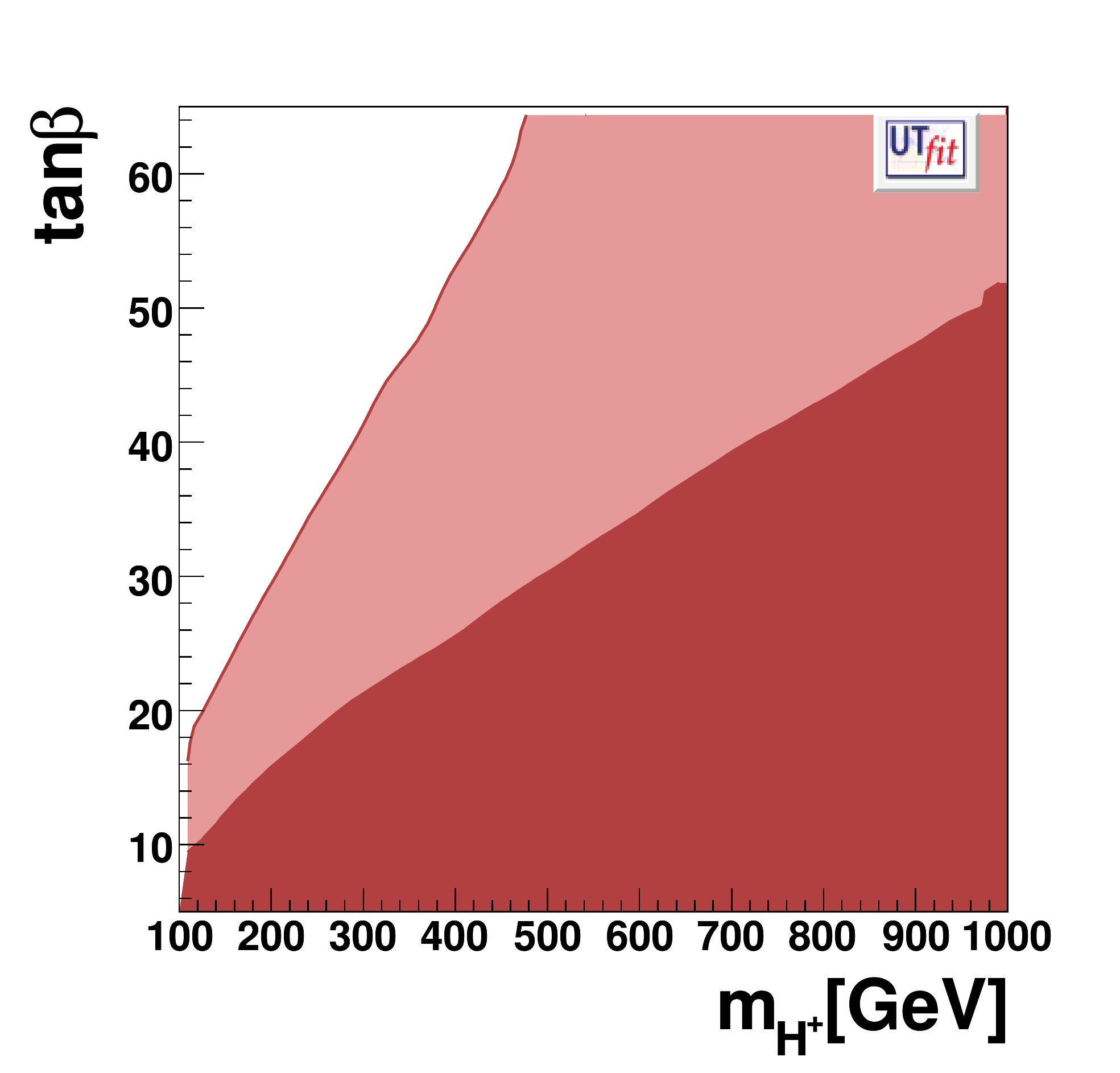}
 \includegraphics[width=0.22\textwidth]{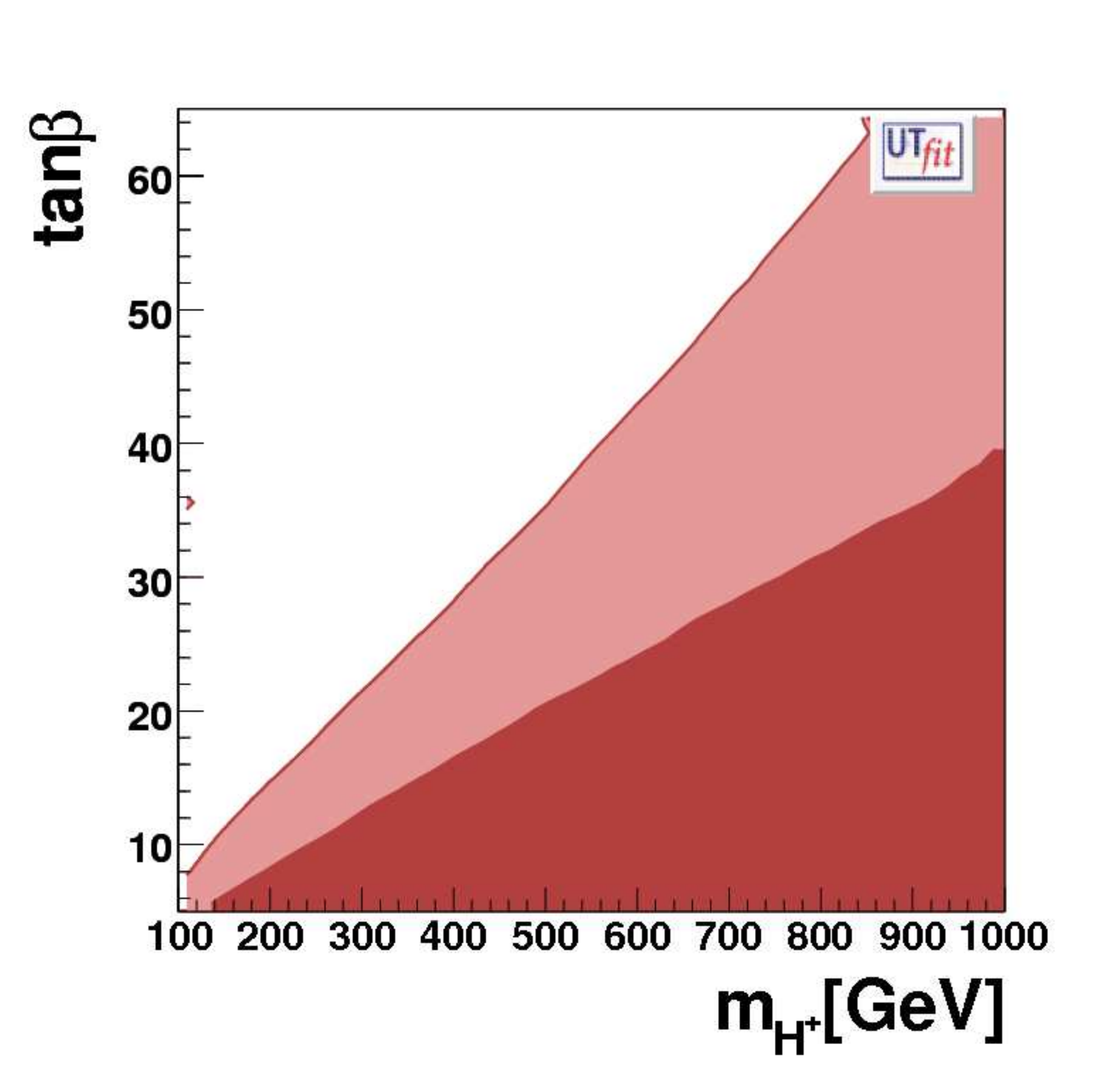}
 \caption{$68\%$ (dark) and $95\%$ (light) probability regions in the
 ($m_{H^+},\,\tan\beta$) plane obtained using $BR(B\to\tau\nu)$ (top left),
 $BR(B_s\to\mu^+\mu^-)$ (top right), $\Delta m_s$ (bottom left), 
 all constraints (bottom right) for $\mu >0$ in the considered MFV-MSSM
 for the parameter ranges specified in the text.\label{fig:MSSMp}}
\end{figure}

\begin{figure}[htb]
 \centering
 \includegraphics[width=0.22\textwidth]{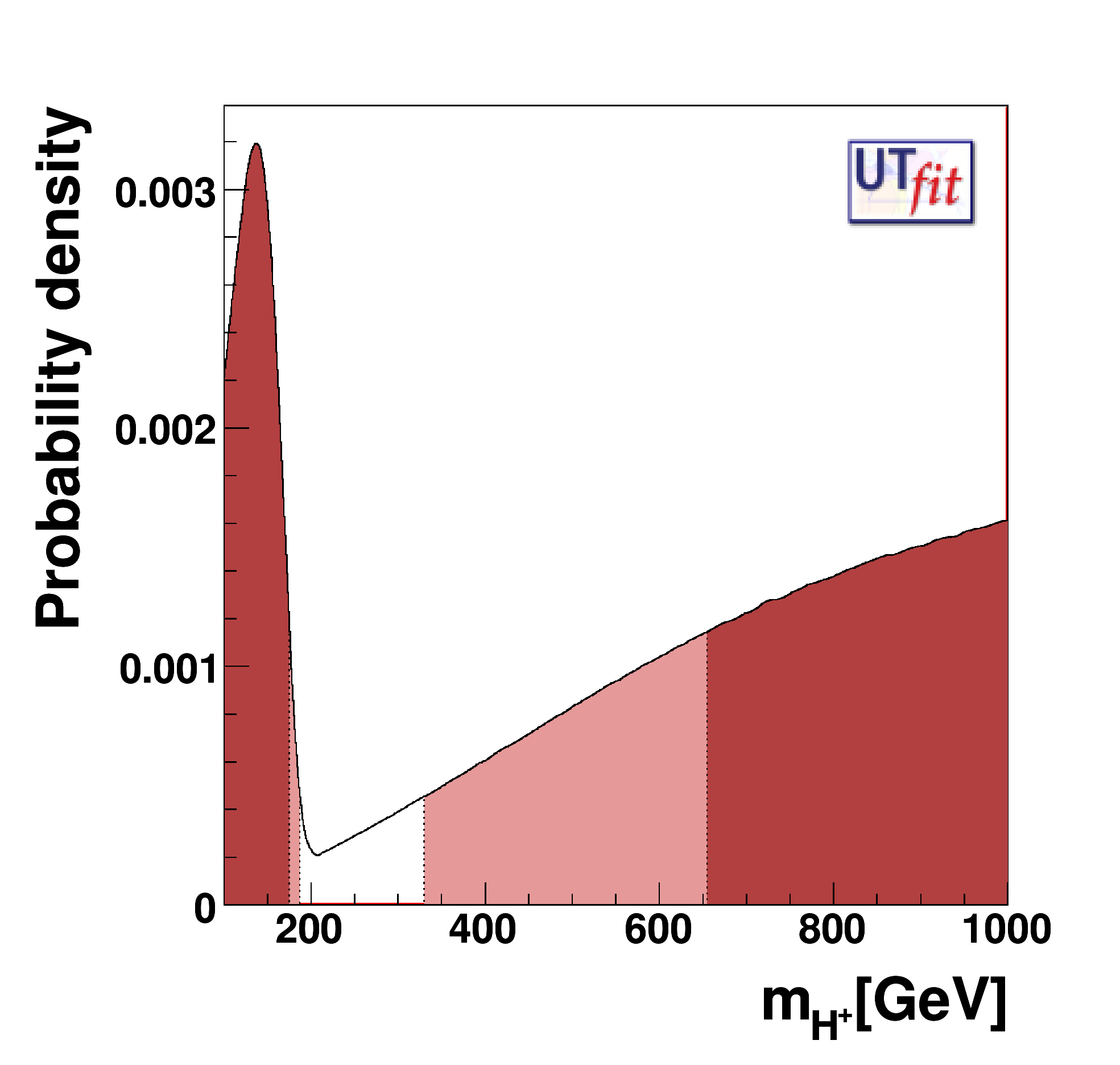}
 \includegraphics[width=0.22\textwidth]{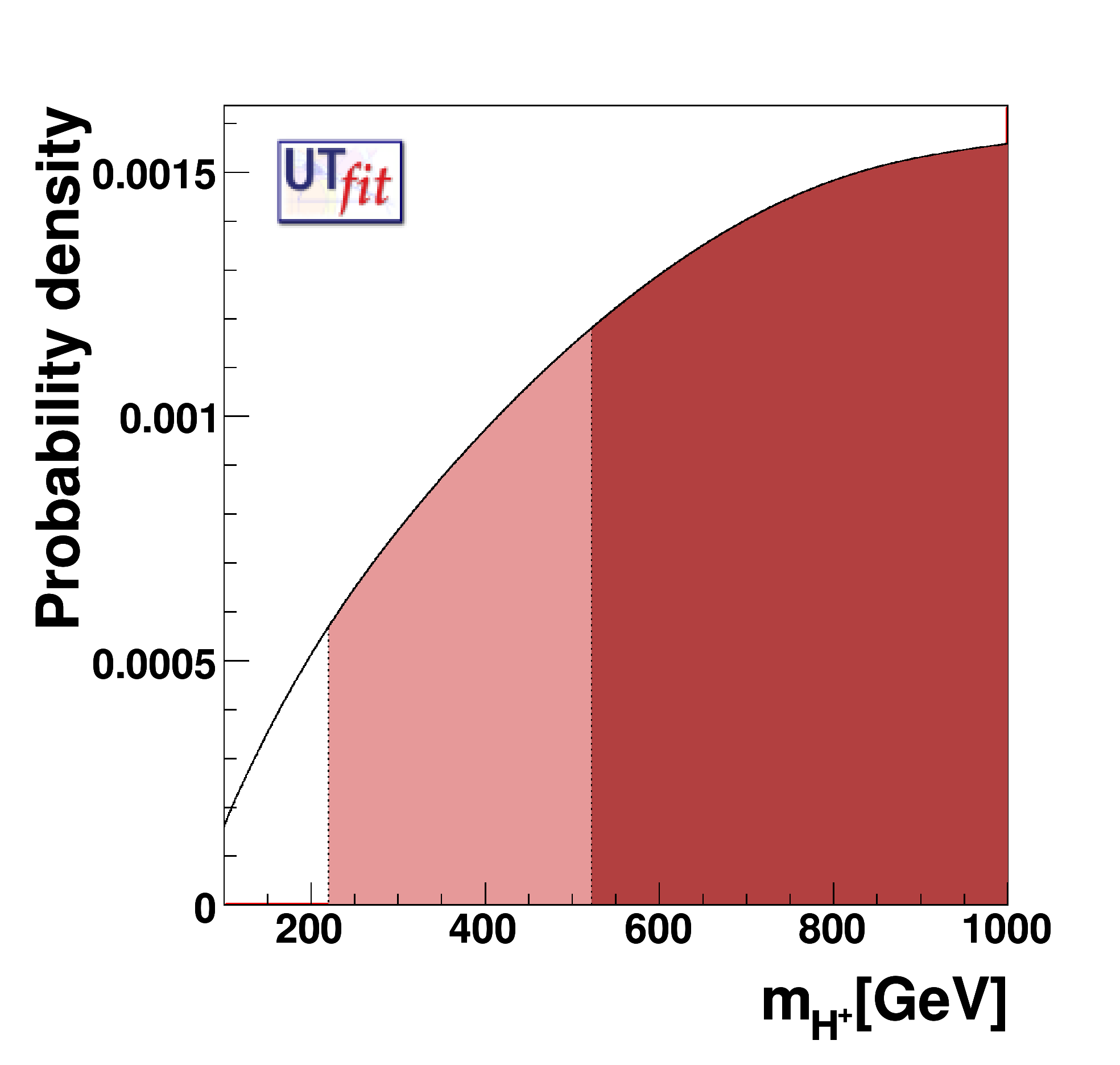}
 \includegraphics[width=0.22\textwidth]{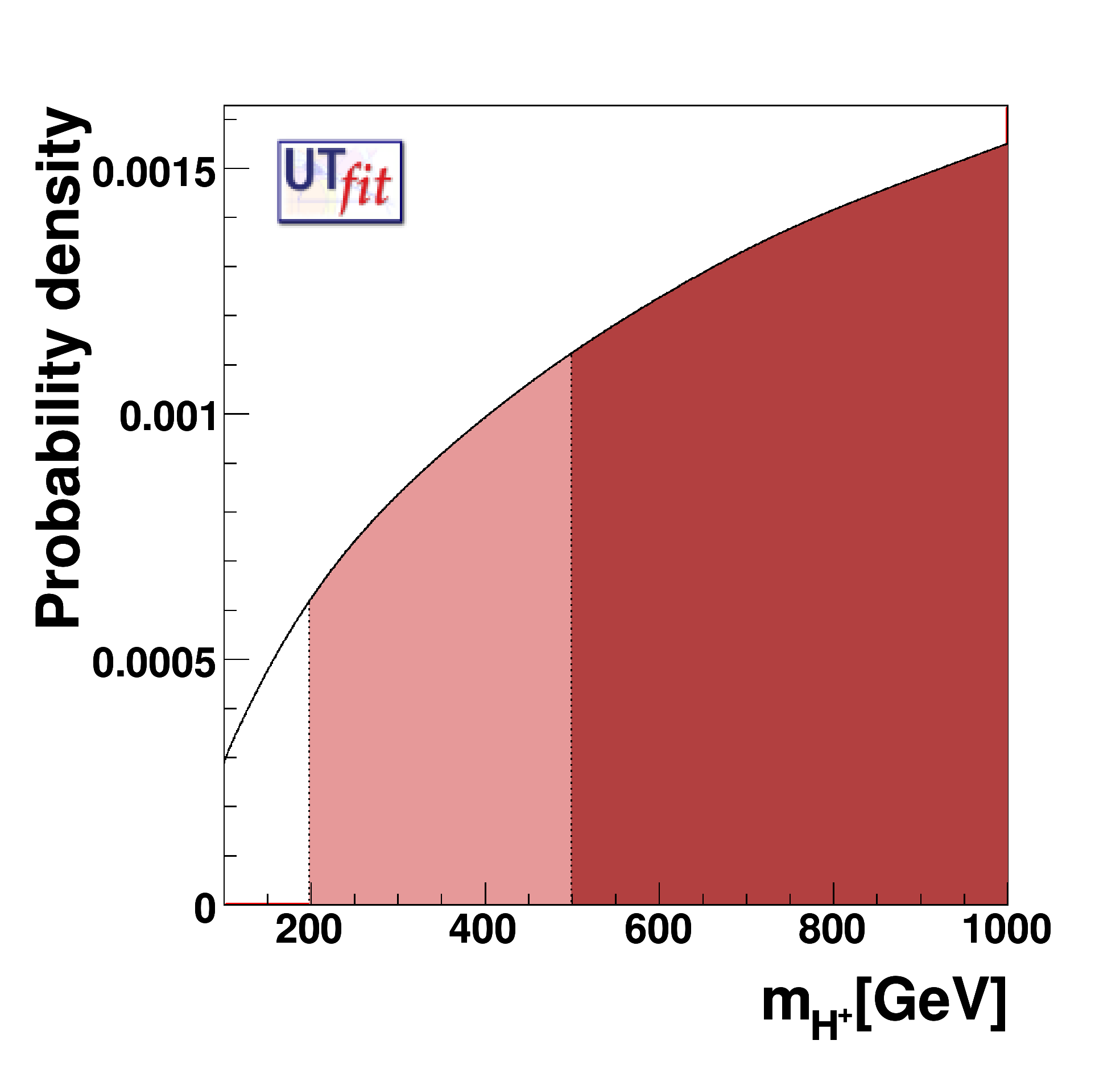}
 \includegraphics[width=0.22\textwidth]{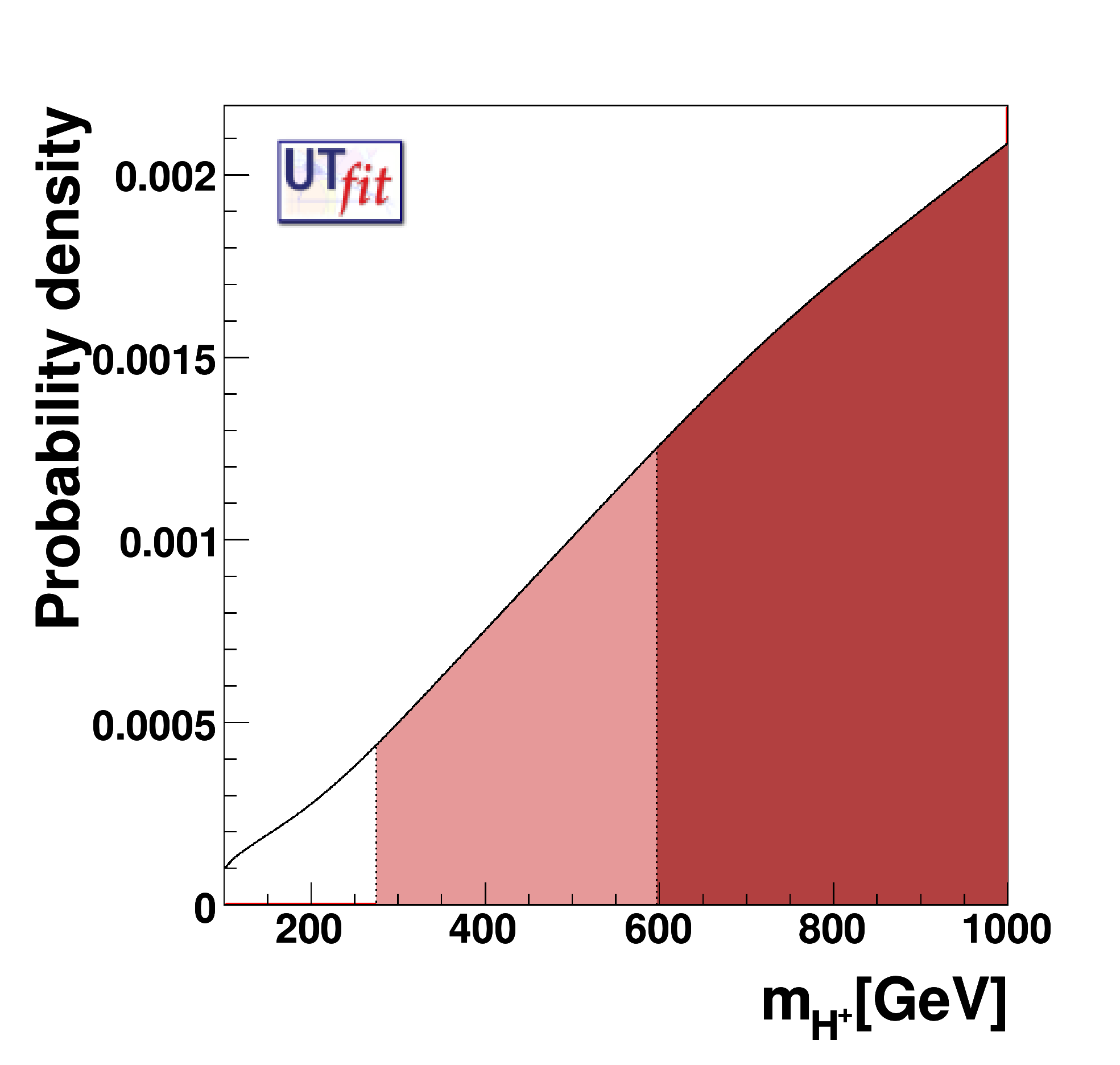}
 \caption{$68\%$ (dark) and $95\%$ (light) probability regions for
 $m_{H^+}$ obtained using $BR(B\to\tau\nu)$ (top left),
 $BR(B_s\to\mu^+\mu^-)$ (top right), $\Delta m_s$ (bottom left), 
 all constraints (bottom right) for $\mu >0$ in the considered MFV-MSSM
 for the parameter ranges specified in the text.\label{fig:MHp}}
\end{figure}

\begin{figure}[htb]
 \centering
 \includegraphics[width=0.22\textwidth]{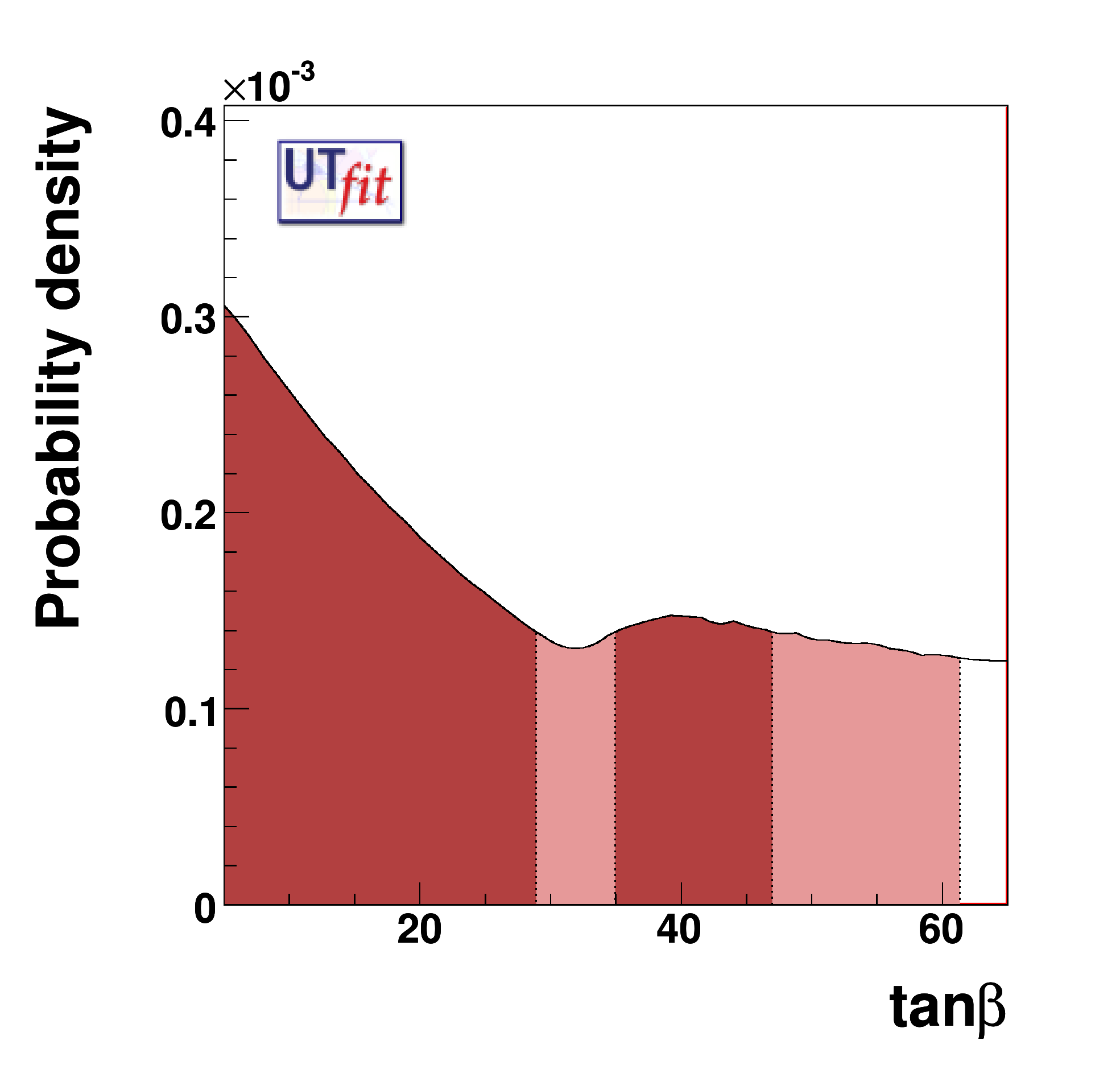}
 \includegraphics[width=0.22\textwidth]{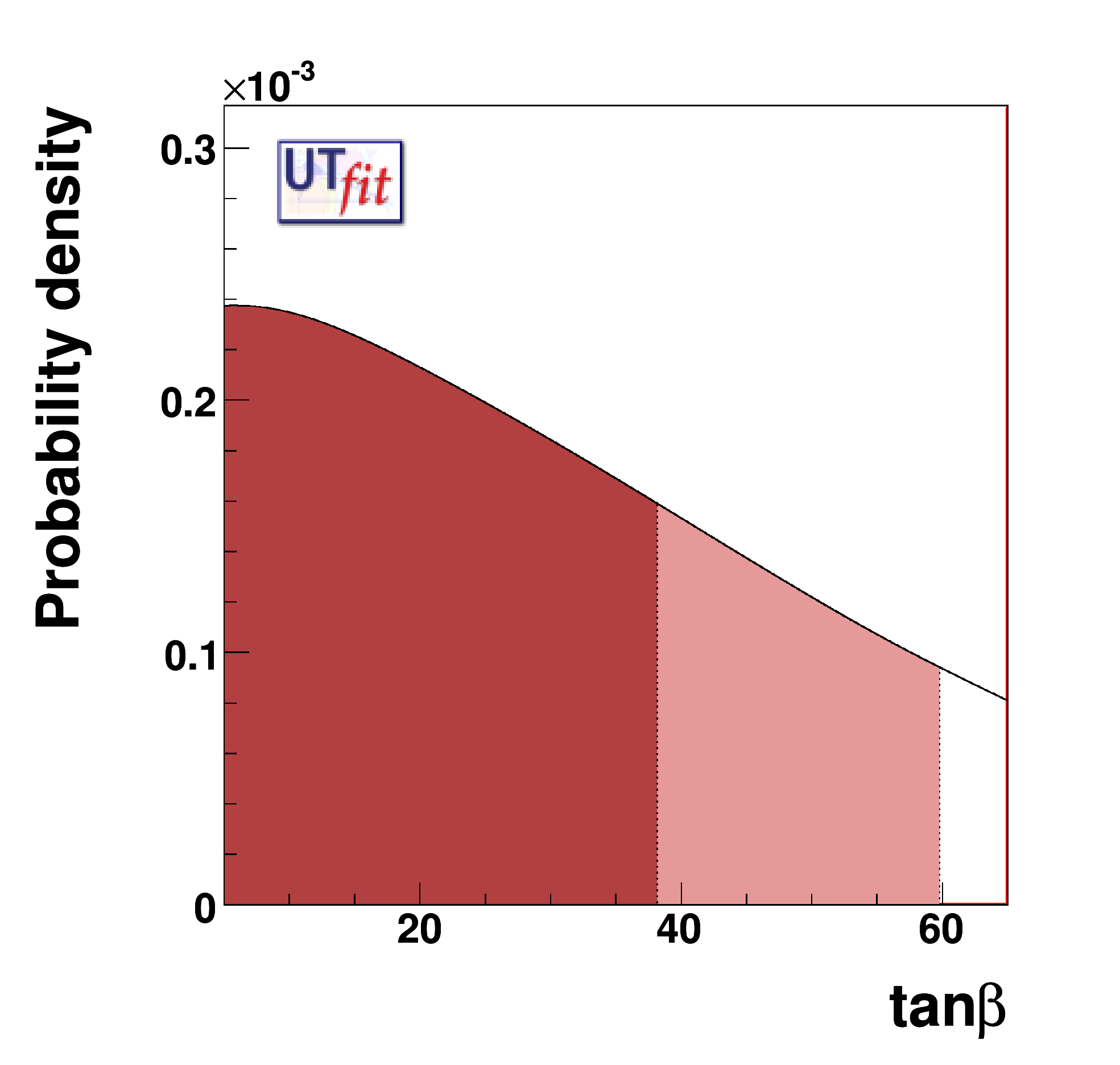}
 \includegraphics[width=0.22\textwidth]{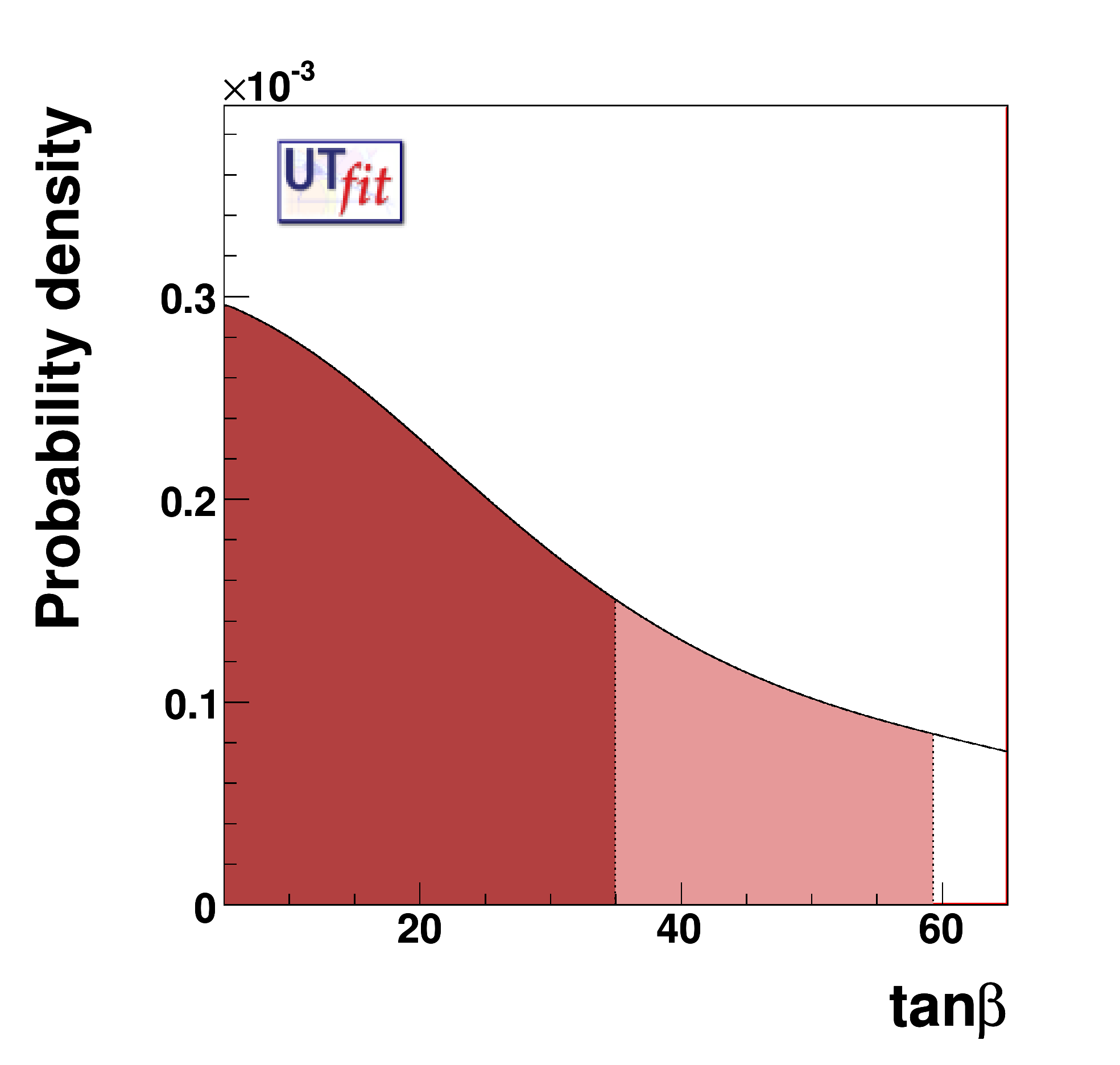}
 \includegraphics[width=0.22\textwidth]{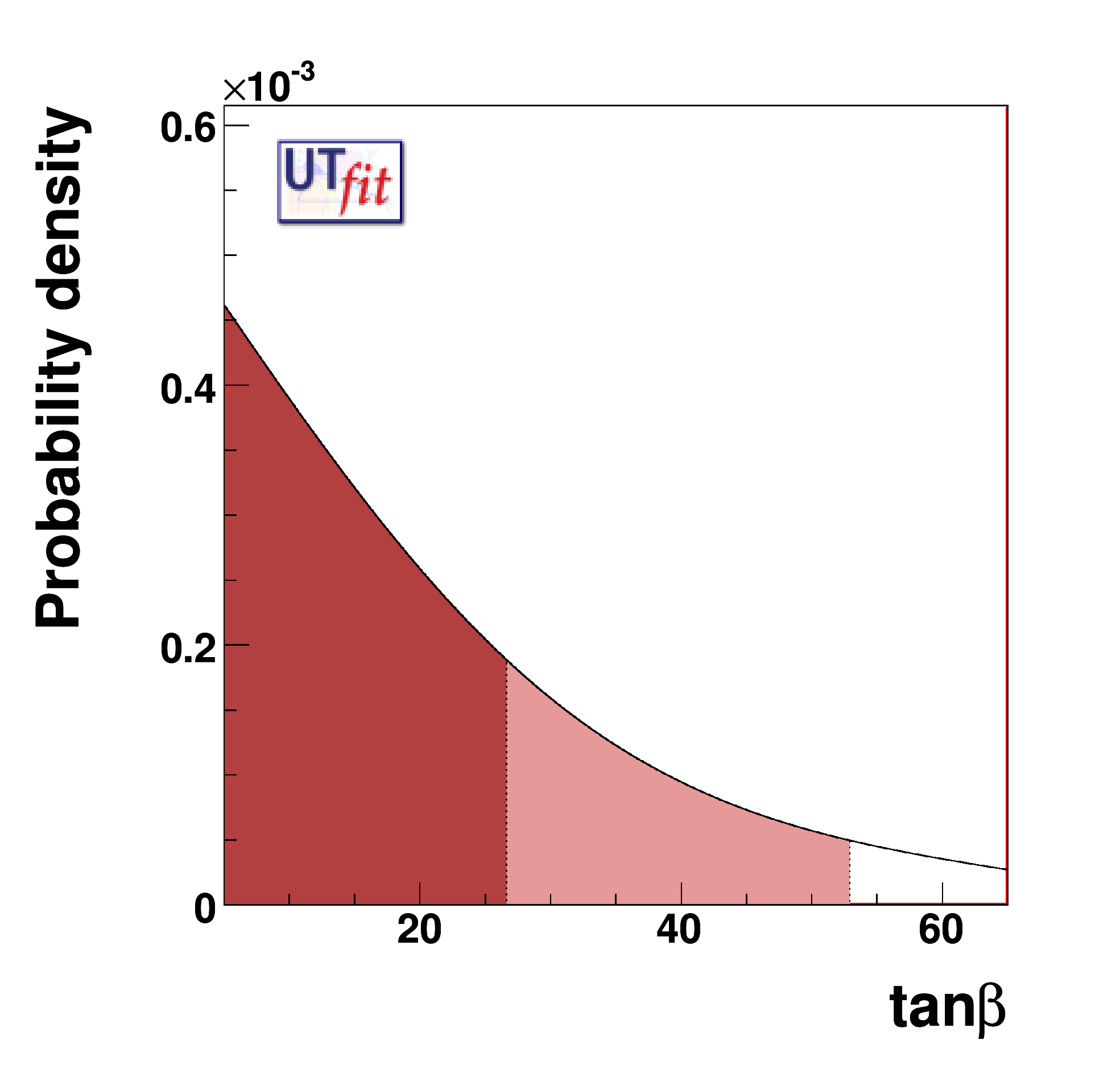}
 \caption{$68\%$ (dark) and $95\%$ (light) probability regions for
 $\tan\beta$ obtained using $BR(B\to\tau\nu)$ (top left),
 $BR(B_s\to\mu^+\mu^-)$ (top right), $\Delta m_s$ (bottom left), 
 all constraints (bottom right) for $\mu >0$ in the considered MFV-MSSM
 for the parameter ranges specified in the text.\label{fig:tanbp}}
\end{figure}

\begin{figure}[htb]
 \centering
 \includegraphics[width=0.22\textwidth]{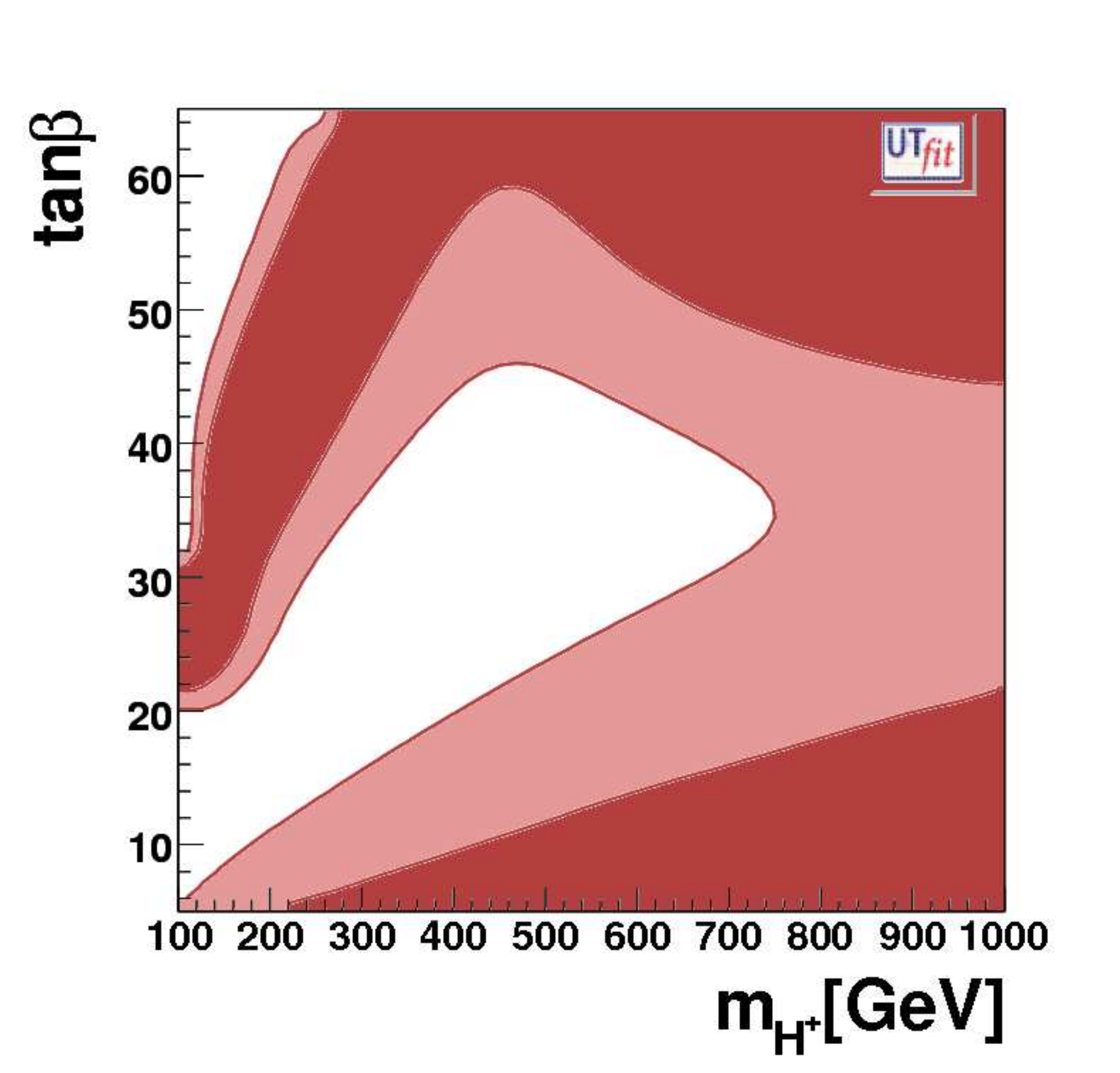}
 \includegraphics[width=0.22\textwidth]{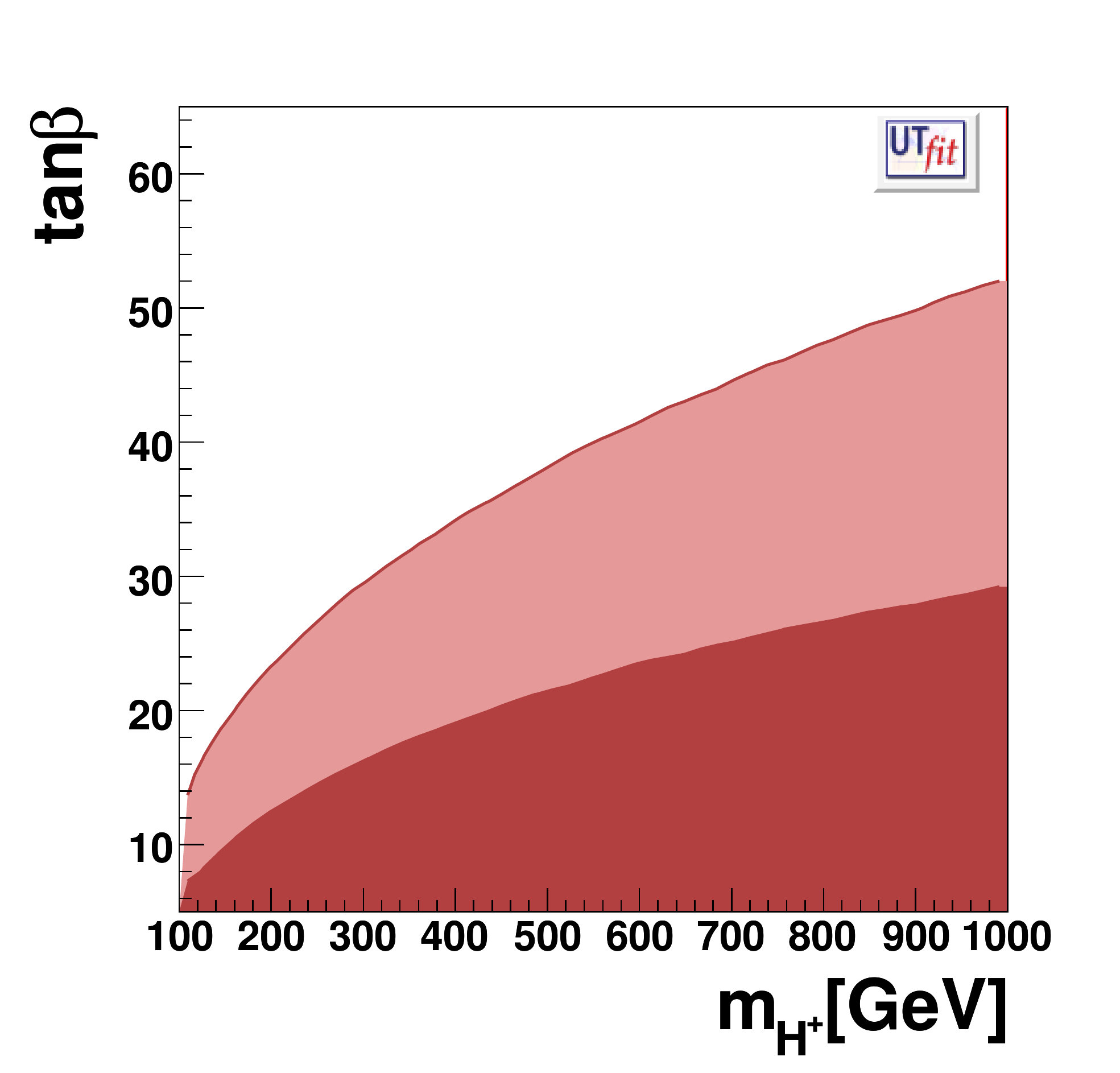}
 \includegraphics[width=0.22\textwidth]{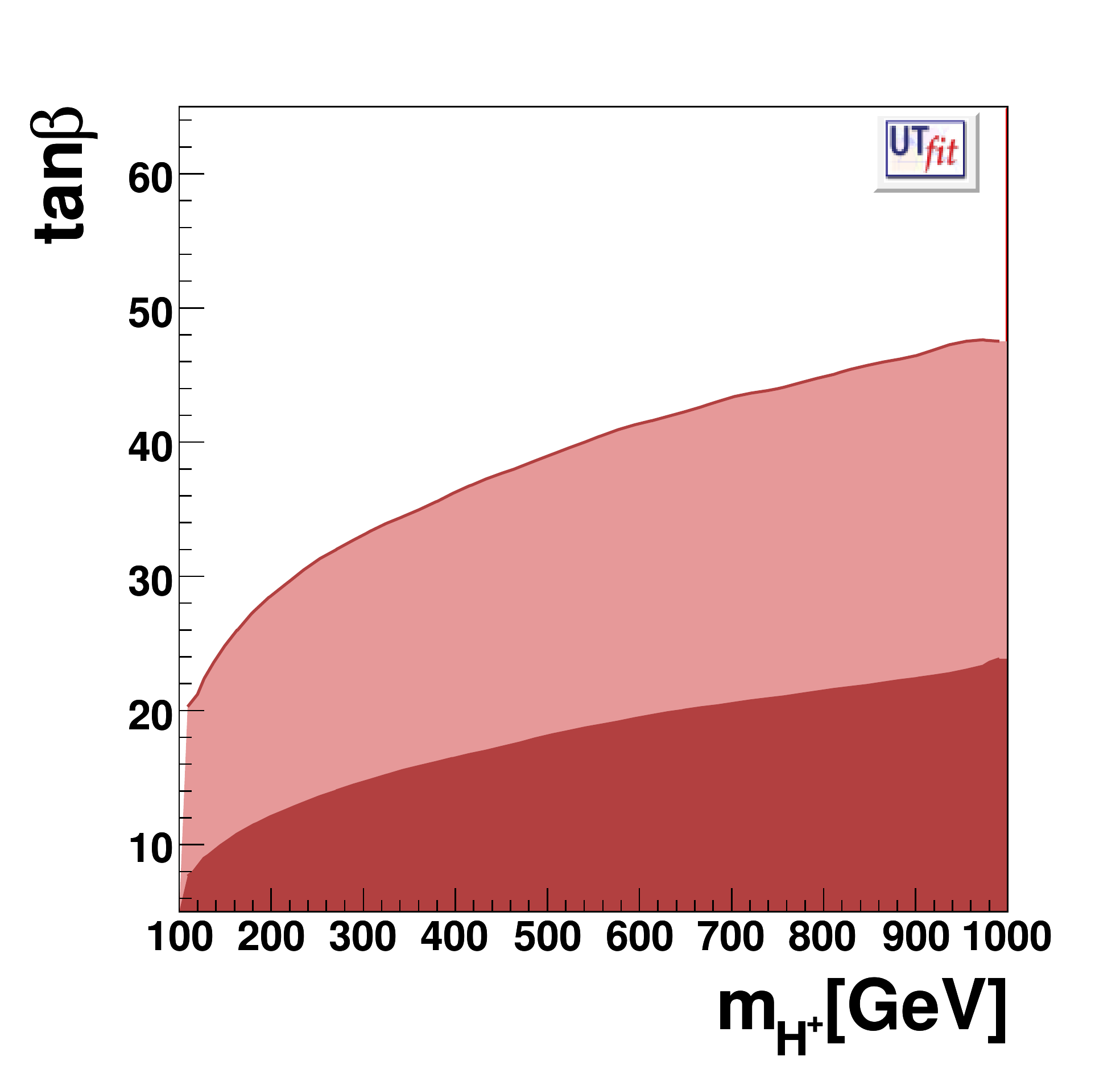}
 \includegraphics[width=0.22\textwidth]{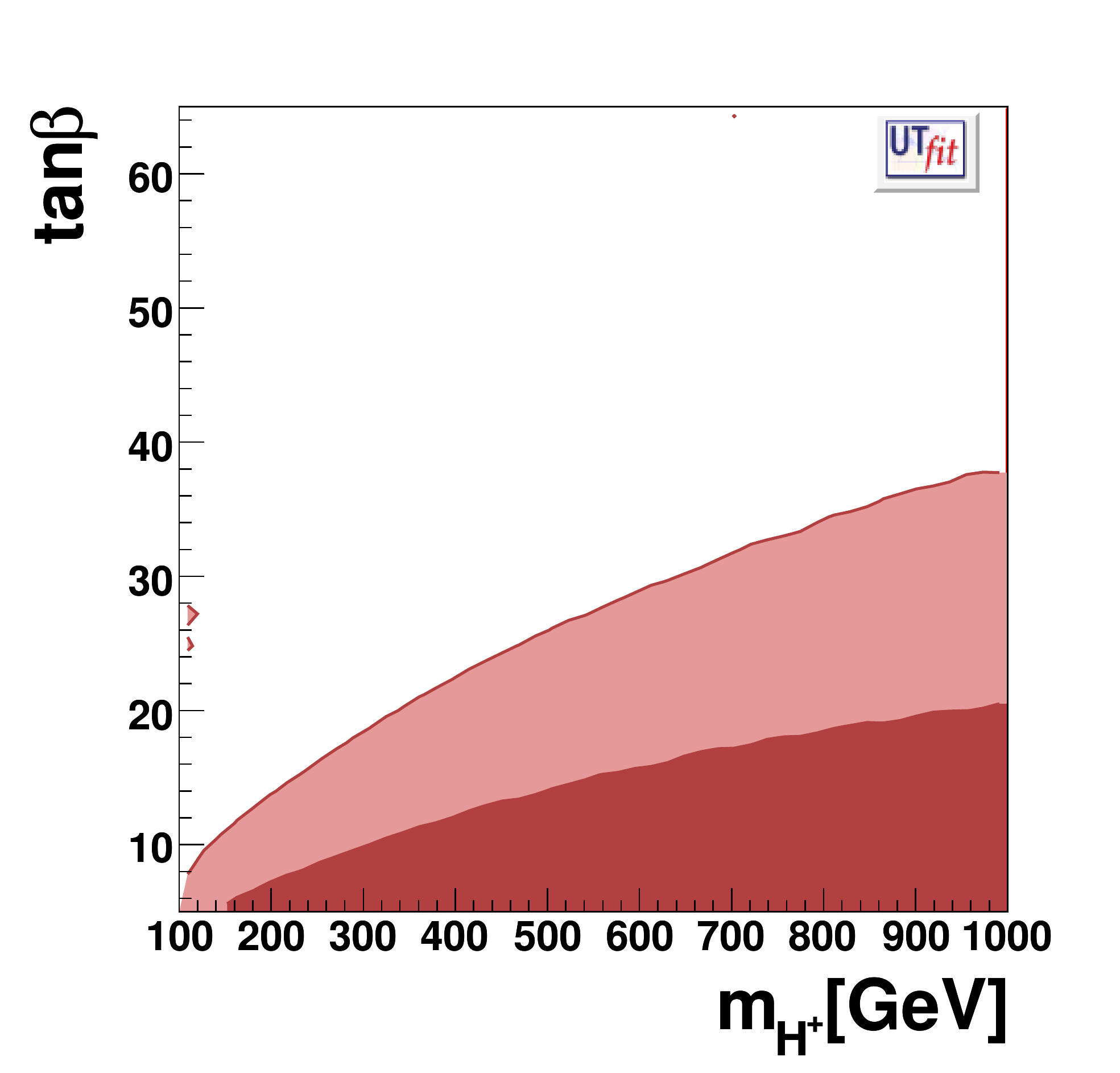}
 \caption{Same as Fig.~\ref{fig:MSSMp} for $\mu<0$.\label{fig:MSSMn}}
\end{figure}

\begin{figure}[htb]
 \centering
 \includegraphics[width=0.22\textwidth]{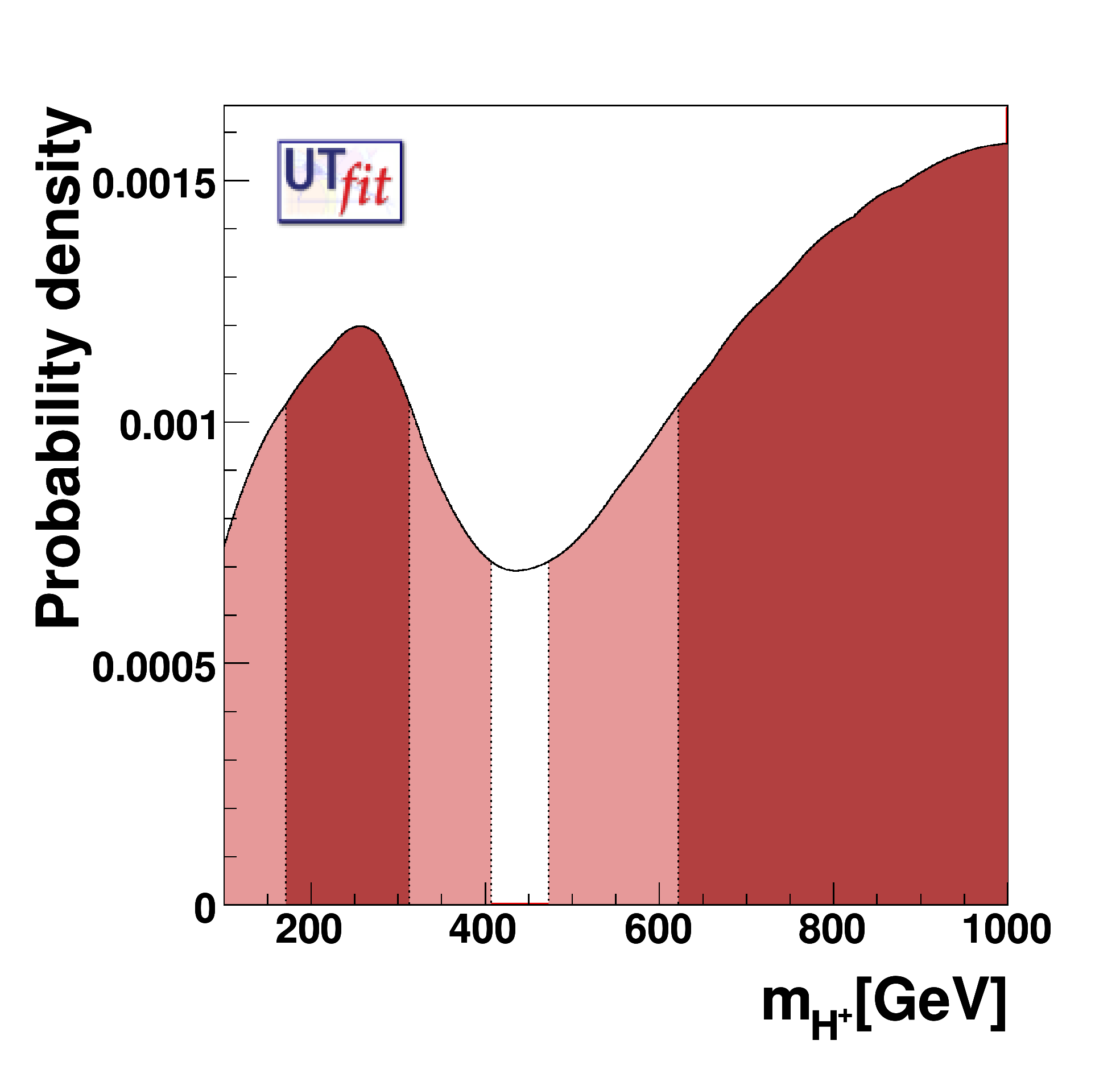}
 \includegraphics[width=0.22\textwidth]{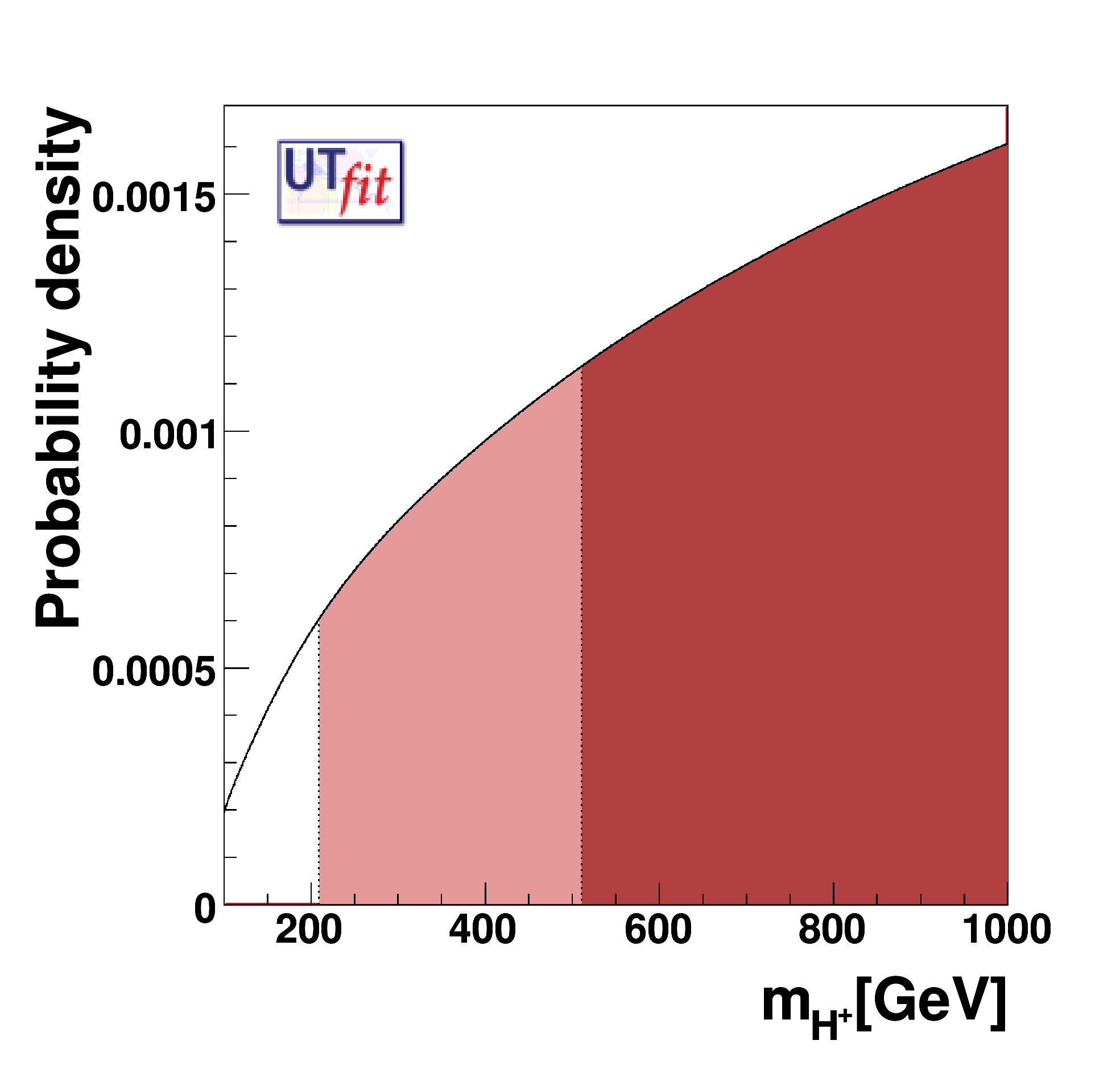}
 \includegraphics[width=0.22\textwidth]{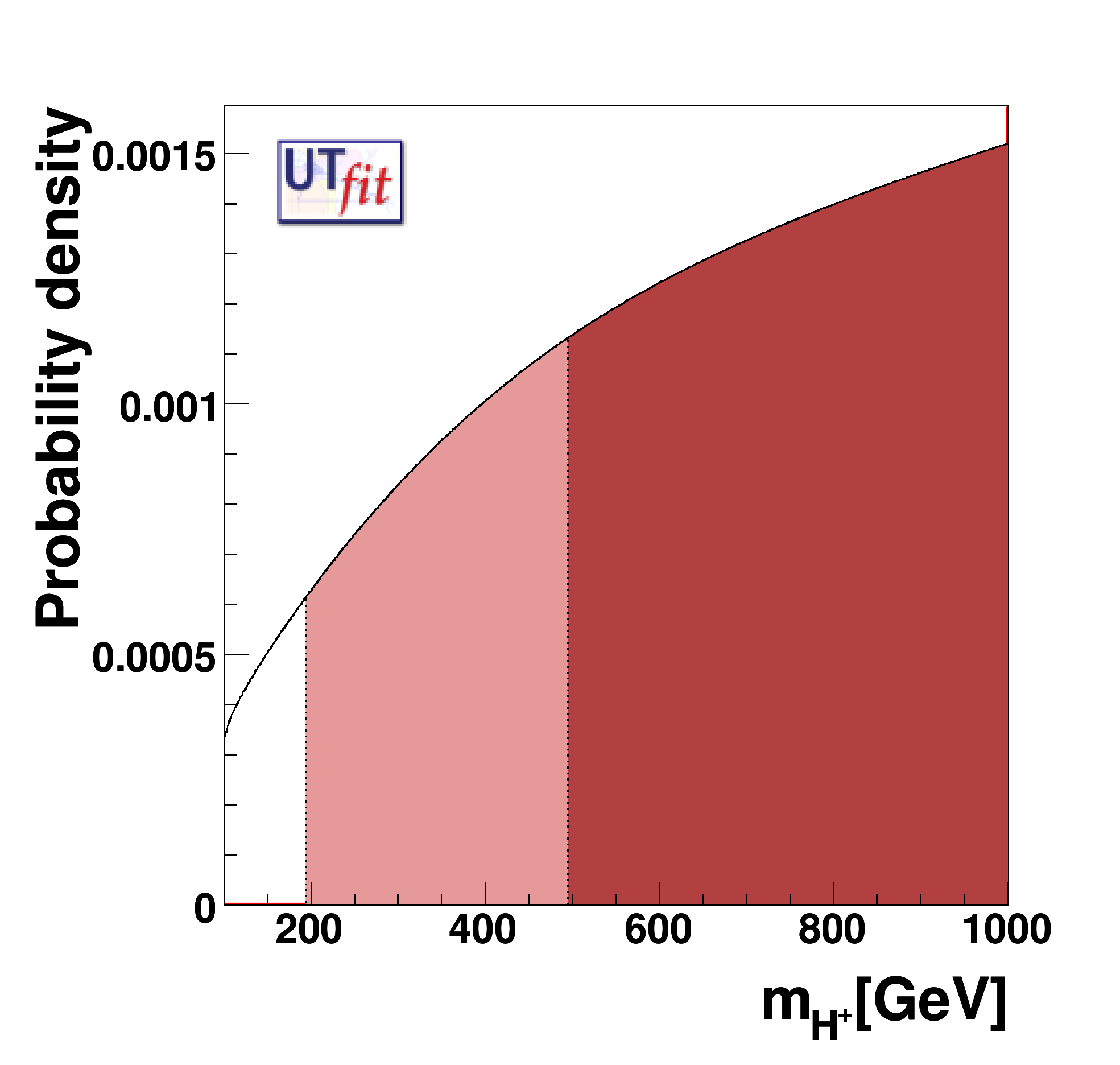}
 \includegraphics[width=0.22\textwidth]{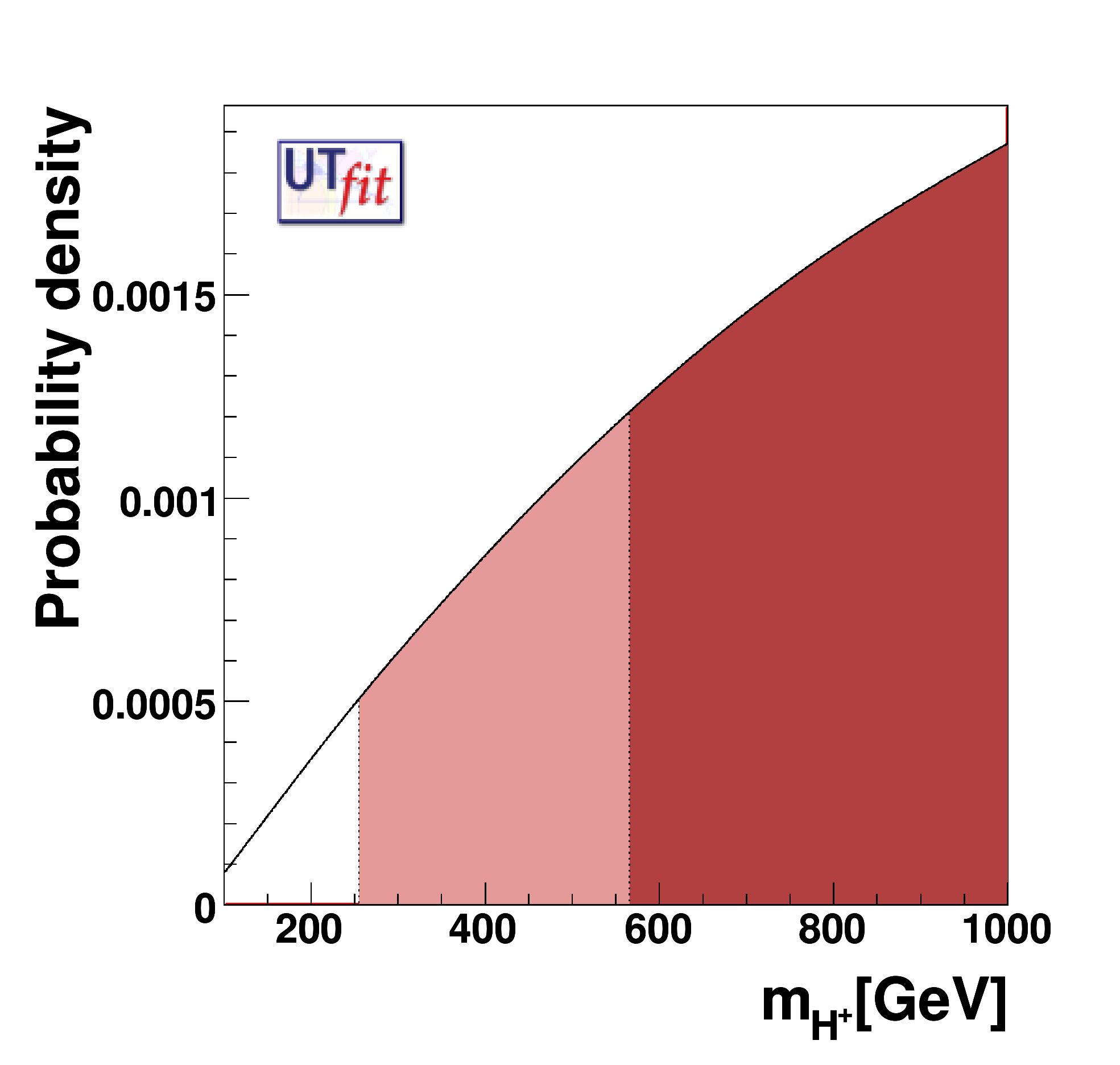}
 \caption{Same as Fig.~\ref{fig:MHp} for $\mu<0$.\label{fig:MHn}}
\end{figure}

\begin{figure}[htb]
 \centering
 \includegraphics[width=0.22\textwidth]{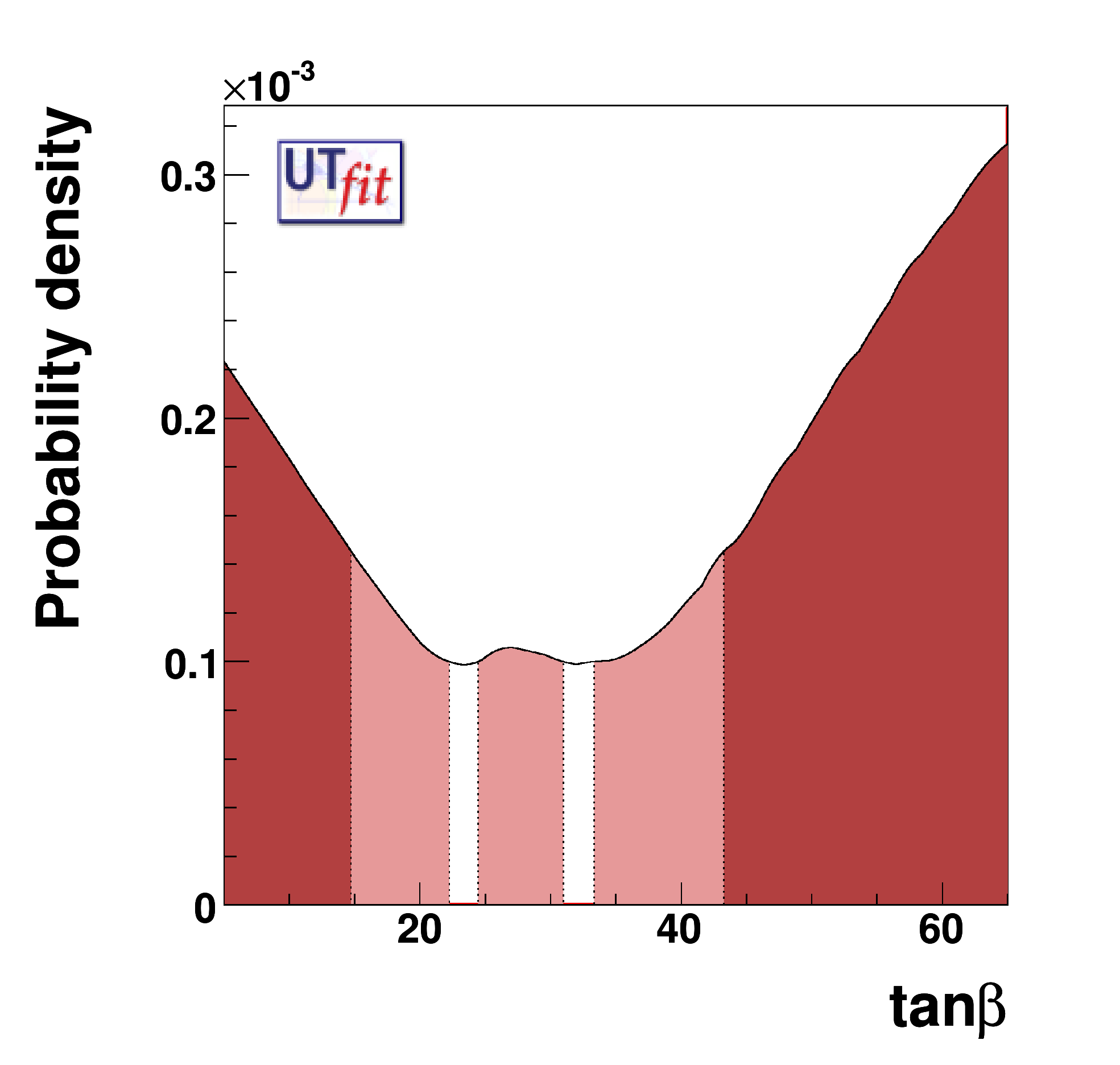}
 \includegraphics[width=0.22\textwidth]{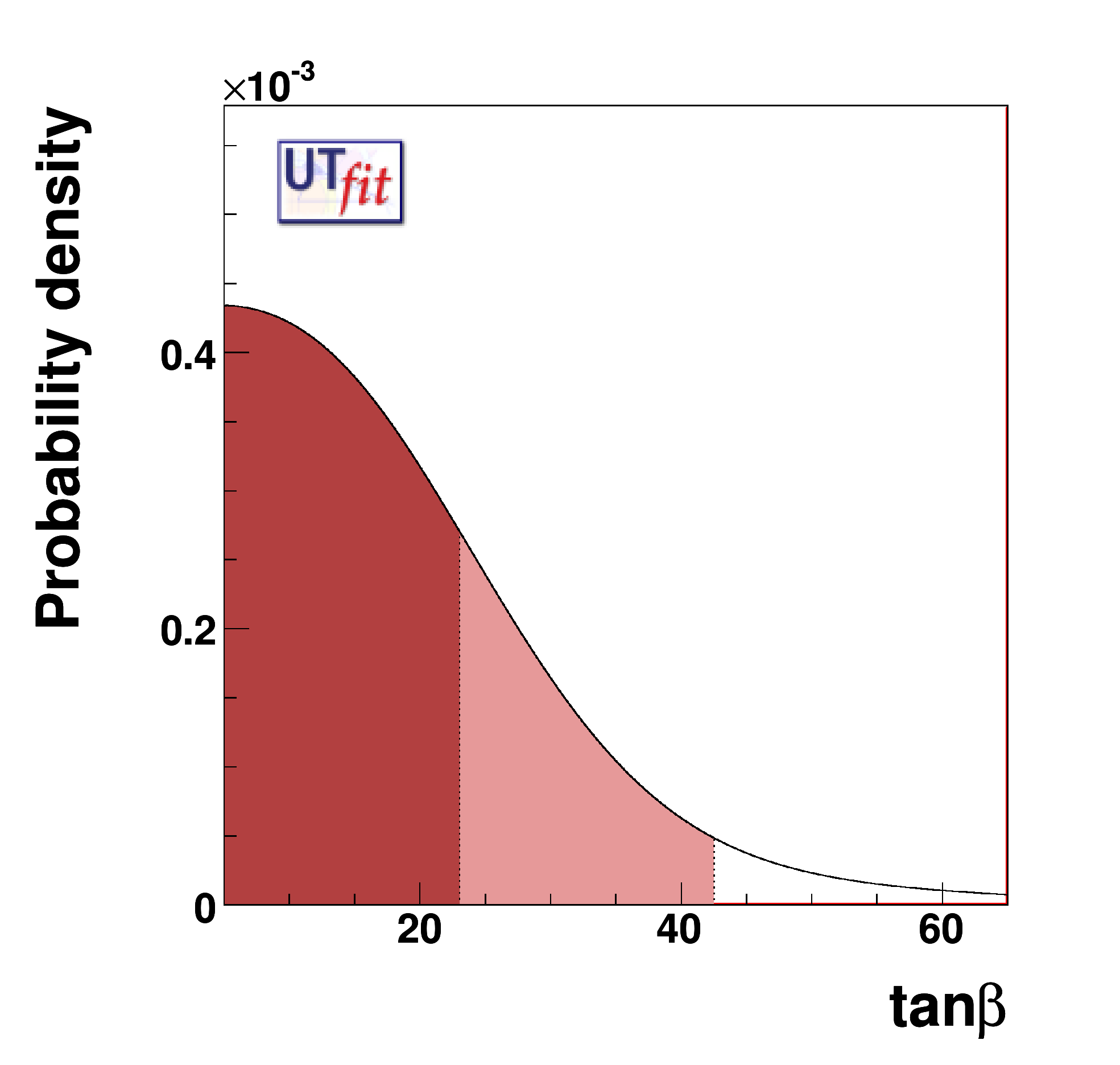}
 \includegraphics[width=0.22\textwidth]{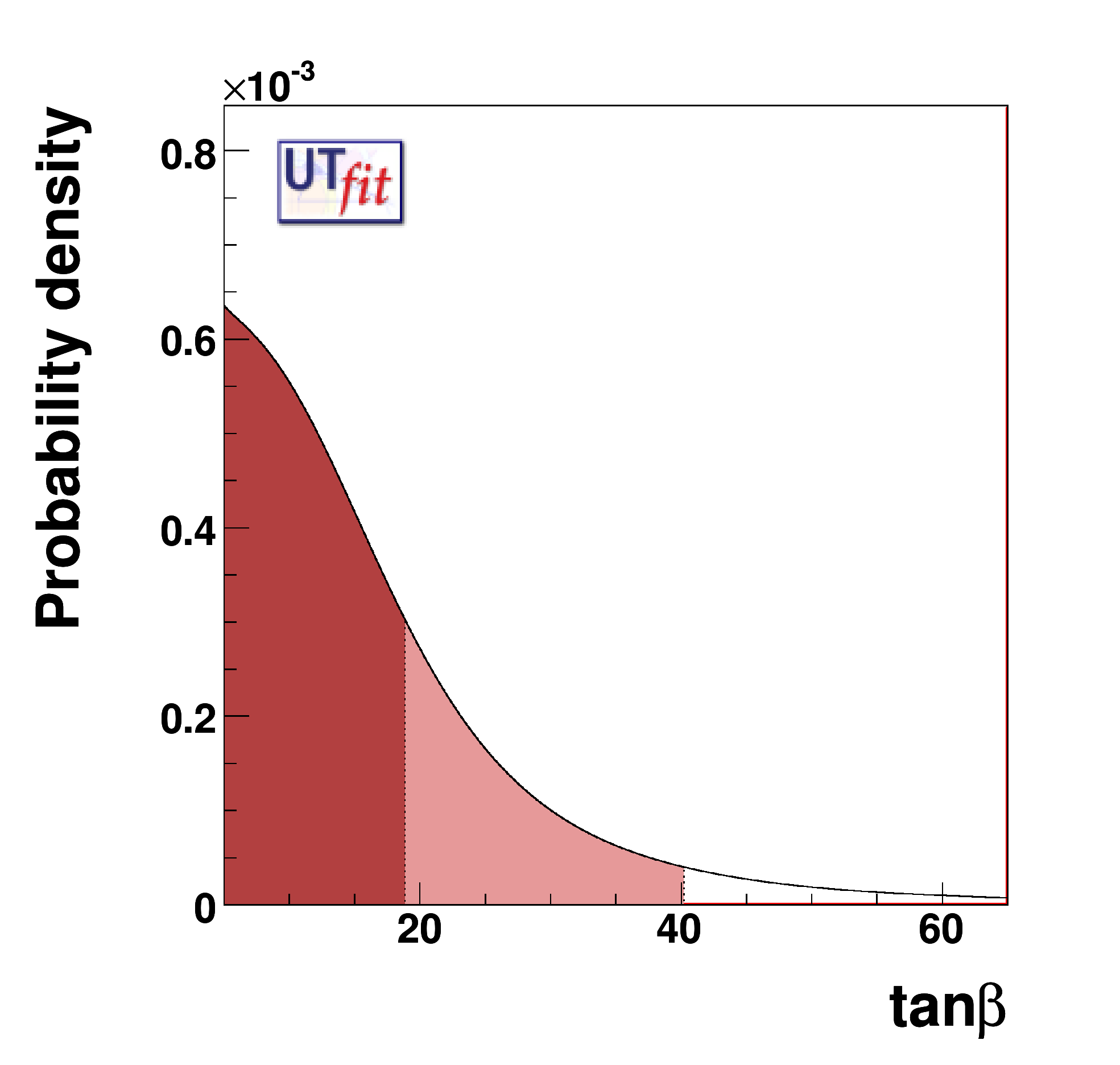}
 \includegraphics[width=0.22\textwidth]{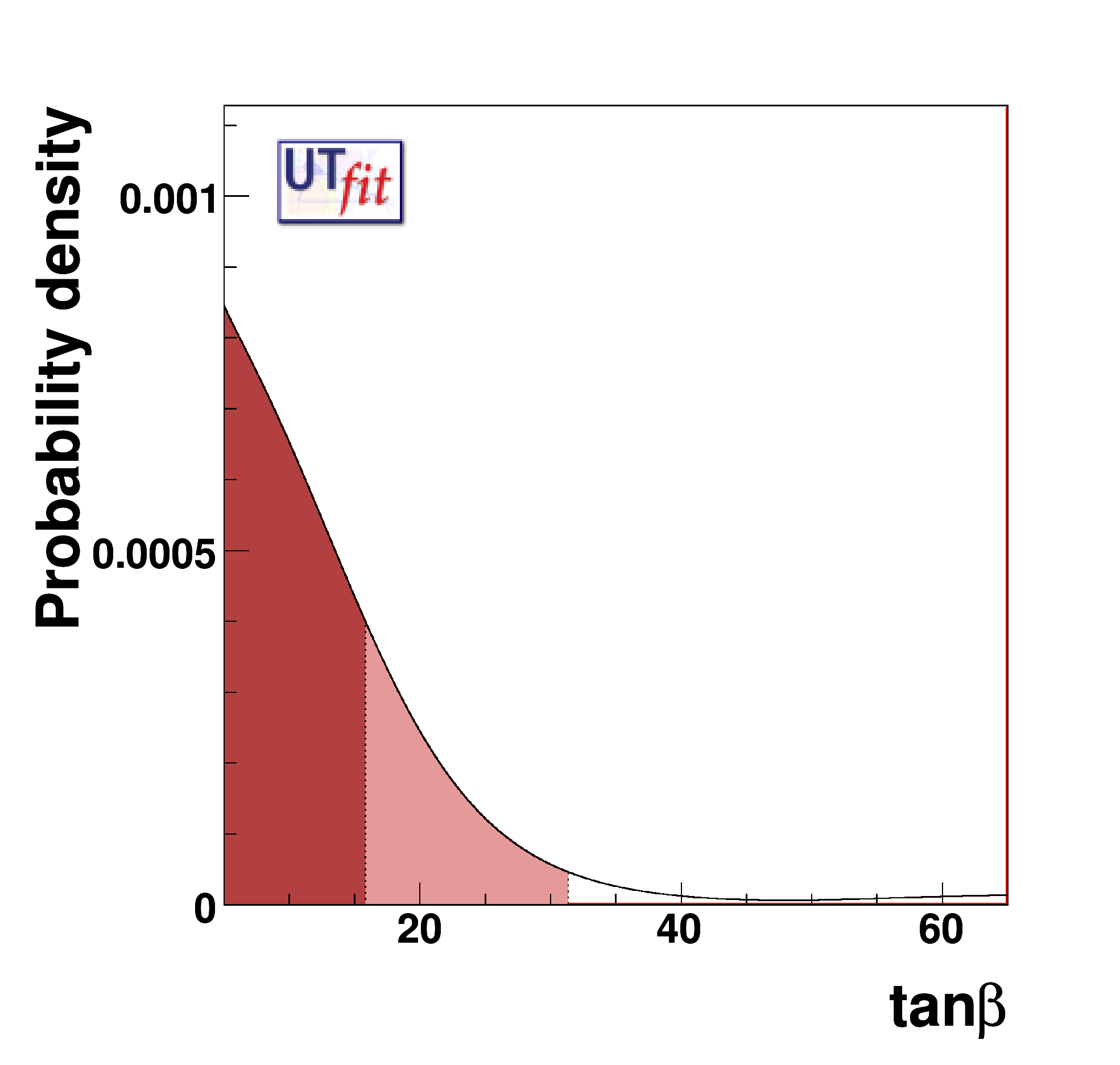}
 \caption{Same as Fig.~\ref{fig:tanbp} for $\mu<0$.\label{fig:tanbn}}
\end{figure}

It has been pointed out that the MSSM with MFV, TeV sparticles and
large $\tan\beta$ could give negligible contributions to flavour
physics except for $B\to \tau\nu$, $\Delta m_s$, $B_s\to\mu^+\mu^-$
and $B\to X_s\gamma$~\cite{Paradise}.  We show that, with present
data, the combination of the first three constraints leaves little
space for large $\tan\beta$. This can be easily understood as this
model typically predicts a suppression of $BR(B\to\tau\nu)$ rather
than the enhancement required by the present measurements.  An
enhancement can be obtained only for very large values of $\tan\beta$
which, however, are disfavoured by the other constraints.

We reanalyze the model of Ref.~\cite{Paradise} with the following
a-priori flat ranges for the relevant low-energy SUSY parameters:
$\mu=[-950,-450]\cup[450,950]$ GeV, $A_u=[-3,3]$ TeV,
$\tan\beta=[5,65]$, $m_{H^+}=[100,1000]$ GeV, $m_{\tilde
  q}=[400,1000]$ GeV, $m_{\tilde g}=[400,1000]$ GeV.  The expressions
of $B\to \tau\nu$, $B_s\to\mu^+\mu^-$ and $\Delta m_s$ can be found in
Eqs.~(3), (11) and (14) of Ref.~\cite{Paradise} respectively.  The
experimental constraints are $\Delta m_s=17.77\pm 0.12$
ps$^{-1}$~\cite{dms} and the upper bound
$BR(B_s\to\mu^+\mu^-)<5.8\times 10^{-8}$ at $95\%$ C.L.~\cite{bsmm}.

In Figs.~\ref{fig:MSSMp} we show the p.d.f.\ in the plane
$(\tan\beta,\,m_{H^+})$ for $\mu > 0$. For completeness, in
Figs.~\ref{fig:MHp} and \ref{fig:tanbp} we present the corresponding
one-dimensional p.d.f.\ for $m_{H^+}$ and $\tan\beta$.  As expected,
the constraint from $B\to\tau\nu$ resembles the one obtained in the
2HDM analysis above (see Fig.~\ref{fig:tanbmh2d}). Once the other
constraints are included, however, the region at large
$\tan\beta/m_{H^+}$ is disfavoured. The combined exclusion region is
roughly bounded by a straight line, giving $\tan\beta < 7.3
\, m_{H^+}/(100$ GeV$)$ at $95\%$ probability, with a remarkable
similarity to the 2HDM-II case. 

For $\mu < 0$, the constraint from
$B\to\tau\nu$ is less stringent for large $\tan\beta$, see
Figs.~\ref{fig:MSSMn}-\ref{fig:tanbn}. In fact, for $\mu < 0$ and very
large $\tan \beta$, the interference with the SM in $B\to\tau\nu$
becomes positive. However the combined bound is more severe than for
$\mu>0$: for $m_{H^+} < 1$ TeV, there is an absolute bound on
$\tan\beta < 38$ with at least $95\%$ probability, while from the
one-dimensional distribution in Fig.~\ref{fig:tanbn} we obtain
$\tan\beta < 32$ at $95\%$ probability.  

For both signs of $\mu$,
large values of $\tan\beta$ for sub-TeV charged Higgses are strongly
disfavoured, including the fine-tuned region where the SUSY
contribution enhances $BR(B\to\tau\nu)$ improving the agreement with
the experimental average.

From our analysis we also derive the following ranges for $BR(B_s \to
\mu^+ \mu^-)$:
\begin{eqnarray}
  \label{eq:bsmmp}
    &&[3,8] \times 10^{-9}~@68\%~\mathrm{prob.}\\
    &&[2,26] \times 10^{-9}~@95\%~\mathrm{prob.}\nonumber
\end{eqnarray}
for $\mu>0$, and 
\begin{eqnarray}
  \label{eq:bsmmm}
  &&[3,6] \times 10^{-9}~@68\%~\mathrm{prob.}\\
  &&[2,17] \times 10^{-9}~@95\%~\mathrm{prob.}\nonumber
\end{eqnarray}
for $\mu<0$.  These ranges can be compared with the SM prediction
$BR(B_s \to \mu^+ \mu^-)_\mathrm{SM}=(3.7 \pm 0.5) \times 10^{-9}$.

\section{Conclusions}
We have shown how the use of the UT fit allows to improve the prediction of
$BR(B \to \tau \nu)$ in the SM, thanks to a better determination of
$\vert V_{ub}\vert$ and $f_B$.
Considering the generalization of the UT fit to various NP scenarios,
we have obtained results for $\overline{BR}$, defined as the
prediction of $BR(B\to\tau\nu)$ assuming negligible NP contributions to
the decay amplitude. The comparison of $\overline{BR}$ to the experimental
result provides an improved probe of the presence of NP in the decay amplitude.
Our results are summarized in Table~\ref{tab:results}.
Finally, we studied the present constraints on the 2HDM-II and on the MFV-MSSM
with TeV sparticles. In both models, we find that large values of $\tan\beta$
for sub-TeV charged Higgs masses are disfavoured by present data.

\begin{acknowledgments}
  We thank G. Isidori for useful comments and discussion on our
  analysis. We acknowledge partial support from RTN European
  contracts MRTN-CT-2006-035482 ``FLAVIAnet'' and MRTN-CT-2006-035505
  ``Heptools''. M.C.{} is associated to the Dipartimento di Fisica,
  Universit\`a di Roma Tre. E.F.{} and L.S.{} are associated to the
  Dipartimento di Fisica, Universit\`a di Roma ``La Sapienza''.
\end{acknowledgments}

\end{document}